%% file: main.tex
\documentclass[trackchanges,twocolumn]{aastex701}

\usepackage{tikz}
\usepackage{multirow}
\usepackage{array}
\usepackage{makecell}
\usepackage{amsmath}
\usepackage{graphicx} 
\usepackage{subcaption}
\usepackage{tabularx}
\usepackage{graphicx}
\usepackage{subcaption}

\usepackage{siunitx}
\usepackage{booktabs}
\usepackage{afterpage}

\usepackage{array}
\usepackage[table]{xcolor}
\newcolumntype{L}[1]{>{\raggedright\arraybackslash}p{#1}}
\newcolumntype{C}[1]{>{\centering\arraybackslash}p{#1}}
\newcolumntype{R}[1]{>{\raggedleft\arraybackslash}p{#1}}
\usetikzlibrary{arrows.meta, positioning, fit, calc, shapes.geometric}

\begin{document}

\title{Asteroids Impacting the Solar System Planets and the Moon. I. Collision Rates from N-body Simulations}

\author[0009-0002-7847-8082]{Qifeng Cheng}
    \affiliation{Department of Physics, Duke University, Durham, NC 27708, USA}
    \email{qifeng.cheng@duke.edu}  
\author[0000-0002-4934-5849]{Daniel Scolnic}
      \affiliation{Department of Physics, Duke University, Durham, NC 27708, USA}
      \affiliation{Department of Electrical and Computer Engineering, Duke University, Durham, NC 27708, USA}
      \email{dan.scolnic@duke.edu}
\correspondingauthor{Qifeng Cheng,
\href{mailto:qifeng.cheng@duke.edu}{qifeng.cheng@duke.edu}}

\begin{abstract}
Previous studies have shown a mismatch between the simulated rate of asteroids impacting the Earth and the observed rate, posing an important problem for planetary defense. Extending this analysis to other planets offers a new opportunity to examine minor body populations and impactor models. We present a unified simulation framework that estimates intrinsic $D\gtrsim10$~m impactor rates on the eight planets and the Moon. We construct size-calibrated source populations of Near Earth Objects (NEOs), Main Belt Asteroids (MBAs), Jupiter-family comets (JFCs), Centaurs, and scattering trans-Neptunian objects (TNOs), and propagate them with a 300-year $N$-body simulation. To avoid relying on black-box-like collision flags, we identify close encounters and estimate impact rates in two complementary ways: a direct count of realized impactors from object-level minimum-distance calculation and a collision expectation based on population-level impact parameter statistics. We predict over 300 years there will be $6.0^{+3.8}_{-2.7}$ impactors on Venus (all NEOs), $10.0^{+11.3}_{-4.9}$ on Earth (NEOs and JFCs), $\sim$$2981^{+481}_{-469}$ on Jupiter (mainly MBAs and JFCs), $1228^{+2759}_{-1017}$ on Saturn (mainly Centaurs), and none on the remaining planets with the $1\sigma$ upper limits spanning $<1.8$ (inner planets) to $<1\times10^{8}$ (outer planets; weak constraint). Jupiter uniquely receives both fast unbound impactors ($>50\, \mathrm{km\,s^{-1}}$) and slower bound ones reaching down to $\sim$$20\,\mathrm{km\,s^{-1}}$. In a companion paper, we compare our predictions with observations and test whether the gap extends across the Solar System and whether current telescopes can observe these collisions.

\end{abstract}

\keywords{\uat{Near-Earth objects}{} --- \uat{Asteroids}{} --- \uat{Solar system dynamics}{} --- \uat{Sky surveys}{} --- \uat{Astronomical simulations}{} --- \uat{Impact processes}{}}
\section{Introduction}
\label{sec:introduction}


The most consequential asteroid impacts are also the ones we are least prepared for, as they occur too rarely for us to routinely observe, test and validate our planetary defense system \citep{Brown2002flux, Brown2013, NRC2010defending}. At the $\sim10$ m size scale, a ``decameter gap" between models and observations emerges \citep{Chow2025decameter}. Fireball and bolide records constrain the meter- to decameter-scale flux at atmospheric entry \citep{Brown2002flux,Silber2009,Brown2013}, yet the inferred decameter impact rate has long disagreed with the rate expected from the survey-debiased NEO population \citep{Brown2013,Harris2021,Deienno2025,NesvornyNEOMOD3}, a tension that has driven recent work on the orbits and physical properties of observed decameter impactors \citep{Chow2025decameter}. Resolving this mismatch tells us whether the dominant uncertainty lies in the source model, the source-to-target delivery dynamics, or the target-dependent detection method. 

This paper and its companion paper address this problem on a Solar System-wide scale. Here (Paper I), we derive intrinsic impact rates across the planets from source populations calibrated to a common size standard. In the companion paper (Paper II) we compare these intrinsic rates with observational estimates from the literature, and forward-model the resulting impactors through survey telescope simulators to include detection biases and derive the detected populations. Together, we close the comparison loop and examine whether the Earth mismatch is target-specific or general across the Solar System. 


One major challenge in such a comparison is the lack of shared conventions in existing simulations and observations. On the source population simulation side, studies typically treat one source-to-target channel in isolation, each with different size conventions. For example, \citet{Chesley2019synthetic} studied the near-Earth object (NEO) population and Earth impactors at decameter-to-kilometer sizes, \citet{ito2010asymmetric} explored the lunar impactors from that same population, and \citet{Nesvorny2023impact} simulated kilometer-scale cometary impactors in the outer Solar System with steady state impact flux estimation on a Gyr timescale. On the target side, impactor observables on each planet are specialized, subject to biases, and inconsistent in size scale. On the Moon, impact flash monitoring probes gram- to kilogram-scale impactors (centimeter-to-decimeter sizes), whereas repeat imaging detects the resulting $\gtrsim 10\,\mathrm{m}$ craters from decimeter-to-meter impactors, with only partly overlapping size ranges \citep{Suggs2014_LunarFlashFlux,liakos2024neliota,speyerer2016quantifying}. On Mars, fresh-crater surveys and seismic detections both track meter-to-decameter craters \citep{DAUBAR2013506,zenhausern2024estimate}. On Jupiter, optical flashes and spacecraft detections probe $\sim$5--20\,m impactors in a high-gravity impact regime, but the inferred rates remain limited by small-number statistics and observing efficiency \citep{2013Hueso,hueso2018small,giles2021detection}. 

A more uniform comparison has recently become possible with the development of improved population models. To quantify survey completeness of the Vera~C.\ Rubin Observatory's Legacy Survey of Space and Time (LSST), a coordinated effort has assembled debiased population models that span the Solar System on a shared computational basis. \citet{Kurlander2025} predict the LSST yield of NEOs, MBAs, and TNOs with the \texttt{Sorcha} survey simulator \citep{merritt2025sorcha}, drawing synthetic populations from debiased models such as NEOMOD3 \citep{NesvornyNEOMOD3} and the Pan-STARRS Synthetic Solar System Model \citep[S3M,][]{Grav2011} for MBAs, and the Canada--France Ecliptic Plane Survey (CFEPS) and the Outer Solar System Origins Survey (OSSOS)-based models for TNOs including scattering-TNOs \citep{Jones2006, Petit2011}. \citet{Murtagh2025} use the same survey simulation pipeline for Centaurs, anchored on OSSOS-calibrated dynamical models \citep{Nesvorny2019}. The steady-state JFC population is based on the calibration of \citet{Nesvorny2017}, whose orbital distribution is traced by the JFC-range particles of the same \citet{Nesvorny2019} integrations. These works provide population models with orbital and color information for reservoirs required here, fitted with the same \texttt{ASSIST}/\texttt{REBOUND} integration tooling \citep{holman2023assist, Rein2012REBOUND, rein2015ias15} that we adopt. Studies of each source population's absolute magnitude distribution, albedo \citep{Ryan2015, Duffard2014, lawler2018ossos}, color \citep{hainaut2012colours, BusBinzel2002, Kurlander2025}, and $H$--$D$ conversion \citep{Harris2002} make our intended consistent size calibration of $\gtrsim 10\,\mathrm{m}$ possible. The adopted magnitude distributions include the diameter-based NEOMOD3 model \citep{NesvornyNEOMOD3}, the binned differential distribution for MBAs \citep{Jedicke2002, Grav2011}, the single power law for Centaurs \citep{Murtagh2025} and for JFCs \citep{Nesvorny2017}, and the divot relation for scattering-TNOs \citep{lawler2018ossos}.

Turning orbital information from the population model into an impact rate can be done in two broad ways. The analytic approach, beginning with \citet{opik1951} and \citet{wetherill1967} and refined by later work \citep{greenberg1982, pokorny2013, JeongAhn2015}, computes a mean collision probability from orbital geometry. It is computationally cheap and underlies most existing cratering- and impact-rate estimates. However, it assumes orbits that are fixed or precess uniformly and encounters that are independent of each other, and its formulae break down for tangential or coplanar geometries \citep{rickman2014}, making this method weakest in the giant-planet regime which our work includes. The second approach records collisions directly in an N-body integration \citep{ito2010asymmetric, NesvornyNEOMOD3}. This is accurate but computationally costly, since impacts are rare and capturing them requires integrating large populations over long timescales, up to $\sim$Gyr for the slowly delivered outer-solar-system sources \citep{Nesvorny2023impact}. And neither approach, when commonly applied, retains the per-encounter geometry that tells one dynamical source-to-target pathway from another, for example, distinguishing a body delivered through a resonance from one scattered inward by the giant planets. 

The two approaches answer two separate questions: how to evolve the orbits and how to score an encounter. We combine them in a two-stage search for impactors. First we flag every object entering a planet's Hill radius (the region where the planet's gravity outweighs the Sun's gravity), and we classify impactors using a minimum-distance (periapsis-distance) estimation and a statistical B-plane impact-probability method \citep{farnocchia2019bplane}. The Hill-radius close encounter criterion provides the link between the two stages, and it is supported by collision-rate studies that sample within a planet's Hill sphere \citep{rickman2014} and mirrors operational impact monitoring on Earth, which tracks close-approach objects across an orbit's uncertainty region \citep{valsecchi2003, milani2005, farnocchia2019bplane, Fuentes-Munoz_2023}. By recovering impact probabilities from these close encounters, the statistical estimators compensate for the sparse direct impact counts produced by the shortened, computationally cheaper integrations. 

The paper is organized as follows. Section~\ref{sec:method} describes the source-population construction, size calibration, Hill-sphere encounter search, periapsis distance impact classifier, and B-plane probability estimator. Section~\ref{sec:results} presents the impact rates, source--target contribution maps, close encounters and impactor dynamics, and time evolution of the simulation. Section~\ref{sec:discussion} interprets the rates in terms of rare-event statistics and uncertainties, examines source-to-target pathways across the Solar System, and discusses the dominant limitations of our work. Section~\ref{sec:conclusion} summarizes the main findings and describes how these intrinsic impactor populations will be used in survey-forward-modeled predictions.

\section{Methodology}
\label{sec:method}
\input{input_population_sum}

\subsection{Input Population and Size Calibration Standard}
To compare impact rates across reservoirs that span very different size and brightness regimes, we calibrate every input population to a common lower size limit. We summarize the resulting populations, with their size and absolute-magnitude thresholds in Table~\ref{tab:population_summary}.

\paragraph{Source populations} We build the input population from five small-body reservoirs including near-Earth objects (NEOs), main-belt asteroids (MBAs), Jupiter-family comets (JFCs), Centaurs, and scattering trans-Neptunian objects (TNOs).

\paragraph{Survey-facing versus calibrated populations}

For each reservoir, except for NEOs and JFCs, we adopt two versions of each population: a survey-facing population and a calibrated (intrinsic) population. For NEOs the two are the same population, and for JFCs we construct only a calibrated population. The survey-facing populations are adopted from existing synthetic catalogs built for survey-completeness and discoverability studies, in particular the LSST-oriented populations of \citet{Kurlander2025} for NEOs, MBAs and scattering-TNOs and \citet{Murtagh2025} for Centaurs. These catalogs already encode photometric assumptions, colors, and detectability cuts, making them well suited for forward-modeling survey detections. The calibrated populations follow the same generation method but omit detectability-related cut and extend to a shared lower size limit. 

\paragraph{Size standard} We anchor the calibrated populations to a common lower size limit of $\gtrsim 10\,\mathrm{m}$. The NEO model (NEOMOD3; \citealt{NesvornyNEOMOD3}) is defined in diameter, so we impose this limit without conversion. The models of the other four reservoirs are defined in absolute magnitude of either the $V$-band magnitude $H_V$ or its variant $H_r$, and therefore require conversion. For a given albedo $p_V$, the $H$--$D$ relation \citep{Harris2002}
\begin{equation}
    D = \frac{1329~\mathrm{km}}{\sqrt{p_V}} \times 10^{-H_V/5}
\label{eq:h_to_d}
\end{equation}
converts the $10\,\mathrm{m}$ diameter limit to an $H_V$ limit. Where $H_r$ is the native variable, we then convert $H_V$ to $H_r$ using the color transformation
\begin{equation}
     H_r \simeq H_V - (V-r)
     \label{eq:hv_to_hr}
\end{equation}
for a given $V-r$ color.

Specifically, MBAs and JFCs are sampled in $H_V$ and Centaurs and scattering-TNOs are sampled in $H_r$. Assuming a survey-consistent MBA albedo $p_V \simeq 0.134$ \citep{Ryan2015}, the $10\,\mathrm{m}$ limit corresponds to $H_V \simeq 27.8$. For JFCs we adopt $p_V = 0.04$, consistent with the albedo assumed in the steady-state population calibration of \citet{Nesvorny2017}, which gives $H_V \simeq 29.11$ at $D = 10\,\mathrm{m}$. For Centaurs, we first convert the $10\,\mathrm{m}$ limit to $H_V \simeq 28.67$ assuming a conservative low-end albedo of $p_V \sim 0.06$ \citep[consistent with the dark tail of the distributions in][]{Duffard2014}, and adopt the Minor Bodies in the Outer Solar System (MBOSS) mean Centaur color $V-R = 0.567 \pm 0.131$ \citep[Table 8,][]{hainaut2012colours} as an approximate proxy for $V-r$. This corresponds to $H_r \simeq 28.10 \pm 0.13$, and we adopt the central value $H_r = 28.10$ for Centaur sampling. Scattering-TNOs are also sampled in $H_r$, but no $V-r$ is available for this population. We therefore anchor Equation~\ref{eq:hv_to_hr} to a reference point, $H_r = 8.66$ at $D \simeq 100~\mathrm{km}$ for an assumed albedo $p = 0.04$ \citep{lawler2018ossos}, which gives
\[
D \simeq 100~\mathrm{km}\,\times 10^{(8.66 - H_r)/5}
\left(\frac{0.04}{p}\right)^{1/2}.
\]
A $10\,\mathrm{m}$ scattering-TNO then corresponds to $H_r \simeq 28.66$. 

Given the substantial uncertainties in albedoes and color transformations, especially for JFCs, Centaurs, and scattering-TNOs, these conversions are not precise size cutoffs and should only be read as approximations.

\subsection{Calibrated Population Generation}
\label{sec:method_calibrated_population}

Each calibrated population is generated using the same four-step procedure outlined below, with population-specific details provided in the corresponding subsections (MBAs in Section~\ref{sec:MBA calibration}, Centaurs in Section~\ref{sec:centaur calibration}, JFCs in Section~\ref{sec:jfc_calibration}, scattering-TNOs in Section~\ref{sec:scattering_population}). 
\begin{enumerate}
    \item We draw each object's absolute magnitude ($H_V$ or $H_r$) from the population's magnitude distribution $N(H)$, which also sets the total number of objects $N$ above the adopted size or absolute magnitude limit.
    \item We assign orbits, typically by cloning and perturbing the semi-major axis, eccentricity, and inclination $(a, e, i)$  from a source orbit pool and either perturbing or randomly drawing the angular elements including argument of perihelion, longitude of ascending node, and mean anomaly $(\omega, \Omega, M)$.
    \item We assign broadband colors, i.e.\ the magnitude differences between survey filters (e.g.\ the LSST $g-r$, $r-i$, $i-z$, and $z-y$ colors), which the survey simulator in Paper~II uses to predict each object's brightness in every filter.
    \item We assign $H,G$ or linear phase-curve parameters, where $H$ is the absolute magnitude drawn in step~1 and $G$ is the slope parameter that sets the shape of the phase curve \citep{Bowell1989}.
\end{enumerate} 

Because the calibrated populations (MBAs, JFCs, Centaurs, scattering-TNOs) are too large to simulate directly, we generate random subsets of synthetic objects, treat them as Monte Carlo realizations of the full $D>10\,m$ populations, and scale all resulting number counts accordingly. 

The distribution comparison between calibrated population and the original survey-facing populations can be found in Appendix~\ref{app:population_comparison}.

\subsubsection{MBAs}
\label{sec:MBA calibration}
Our MBA model follows the S3M framework of \citet{Grav2011} for the size distribution and orbit assignment, and \citet{Kurlander2025} for colors and phase functions. We extend the S3M size distribution to smaller sizes and generate our own Monte Carlo realization down to that limit. Each step is described below.

We draw each object's absolute magnitude $H_V$ independently of its orbit from the debiased differential size--frequency distribution of \citet[][Table~1]{Jedicke2002}, as adopted in the S3M framework \citep{Grav2011}. We use the tabulated differential counts $n(H)$ in $0.5$~mag bins from $H = 5.25$ to $18.25$ and, beyond the final bin, extrapolate linearly in $\log_{10} n(H)$ as
\begin{equation}
n(H) = K \times 10^{\alpha H},
\end{equation}
with slope $\alpha = 0.245$ and normalization $K = 13.38$ taken from the final bin at $H = 18.25$. This extends the \citet{Grav2011} extrapolation to the smaller sizes required here. Our size calibration $H_V \simeq 27.8$ yields a total implied population of $\sim 2.565 \times 10^{8}$ objects.

We then assign orbits by perturbation-based cloning of known large main-belt asteroids, following the S3M procedure \citep{Grav2011}. The source orbit pool consists of $36{,}304$ multi-opposition MBAs with $H_V < 14.5$ extracted from the Minor Planet Center MPCORB catalog, and this magnitude regime is generally treated as observationally complete \citep{Jedicke2002}. For each synthetic object, we select one source orbit at random and apply small uniform perturbations to its elements \citep[][Table~1]{Grav2011}: $\Delta a = \pm 0.01$~au, $\Delta e = \pm 0.01$, $\Delta i = \pm 0.5\arcdeg$, and $\Delta\Omega = \Delta\omega = \Delta M = \pm 1.0\arcdeg$. We require each perturbed orbit to satisfy the MBA boundary constraints of semi-major axis $1.8 \leq a \leq 4.1$~au and perihelion distance $q > 1.3$~au and to have $0 < e < 0.99$.

Colors are assigned following the procedure of \citet{Kurlander2025}. Each object is assigned a taxonomic class, C-type or S-type, with C-type probability of $a / 2\,\mathrm{au} - 1$, following \citet{Schwamb2023}. Because C-type probability increases with semi-major axis, this prescription produces a compositional gradient across the belt, with S-types dominating the inner belt and C-types the outer belt, consistent with observations \citep{GradieTedesco1982, DeMeoCarry2014}. Conditioned on the assigned class, we then draw broadband colors by randomly sampling the empirical SMASS~II visible-wavelength spectra of \citet{BusBinzel2002}. We then assign each object an $H,G$ phase function with slope parameter $G = 0.15$, matching the Minor Planet Center default and the implementation of \citet{Kurlander2025}. 

For the subsequent simulations, a random subset of $10^7$ MBAs is used as a Monte Carlo realization of the full ($H_V < 27.8$; $2.565 \times 10^8$ objects), and thus all resulting number counts are scaled by a factor of $25.65$.

\subsubsection{Centaurs}
\label{sec:centaur calibration}
Our Centaur model follows \citet{Murtagh2025} for magnitudes, orbits, colors; we extend it to our fainter size limit with an independent Monte Carlo realization.

We draw each object's absolute magnitude $H_r$ in the LSST $r$ band, and adopt the single-power-law cumulative absolute-magnitude distribution
\begin{equation}
    N(<H_r) = N_0\,10^{\alpha(H_r-H_0)} ,
    \label{eq:centaur_hr_law}
\end{equation}
with slope $\alpha = 0.42$, consistent with OSSOS-based constraints \citep{lawler2018ossos}, and normalization $N_0 = 21{,}400$, $H_0 = 13.7$, matching the debiased Pan-STARRS1 estimate of \citet{Kurlander2025panstarrs1}. We assign each object an $H_r$ value by inverse-transform sampling of Equation~\ref{eq:centaur_hr_law}. Our size calibration $H_r \simeq 28.1$ yields a total implied population of $\sim2.39\times10^{10}$ objects.


We then assign orbits by sampling the cloned-particle integrations of \citet{Nesvorny2019}. The source orbit pool consists of $\sim25$ million rows---approximately $98\%$ of the original catalog---retained under the Gladman et al. dynamical definition \citep[G08;][]{gladman2008nomenclature}, $a < 30.1~\mathrm{au}$ and $q = a(1-e) > 7.35~\mathrm{au}$, which selects objects interior to Neptune while excluding the most Jupiter-family-comet-like orbits. For each synthetic object, we draw one orbit at random (with replacement) from the source pool and copy its $(a, e, i)$ directly, preserving the correlated orbital structure of the steady-state Centaur population. We randomize only the angular elements, drawing each independently from $\Omega, \omega, M \sim U(0^\circ, 360^\circ)$. This removes artificial clustering in orbital longitude while retaining the dynamical structure encoded in the source integrations.

We then assign each object a bimodal Centaur class, blue (Bienor-like) or red (Pholus-like), with class fractions of $75\%$ and $25\%$ following \citet{Wong2017}, and draw the corresponding LSST colors from Table~2 of \citet{Murtagh2025}. Finally, each Centaur is given a linear phase function with a fixed coefficient $\beta = 0.071~\mathrm{mag\,deg^{-1}}$ uniform across all filters. 

Given the large full $H_r < 28.1$ population ($2.39\times 10^{10}$ objects), we generate a random subset of $2\times10^7$ synthetic Centaurs as a Monte Carlo representation and apply a scaling factor of $1195.04$ for the subsequent results.

\subsubsection{JFCs}
\label{sec:jfc_calibration}

Our JFC model follows the population calibration of \citet{Nesvorny2017} and uses the orbital integrations of \citet{Nesvorny2019}, while extending to our smaller size limit.

We sample JFC absolute magnitudes directly in $H_V$ using the same single-power-law cumulative form as Centaurs (Equation~\ref{eq:centaur_hr_law}) but re-expressed in $H_V$, 
\begin{equation}
    N(<H_V) = N_0 \, 10^{\alpha(H_V - H_{V,0})},
    \label{eq:jfc_hv_law}
\end{equation}
with slope $\alpha = 0.42$ and normalization anchored to $N_0 = 700$ objects with $D > 2\,\mathrm{km}$ and $q < 2.5\,\mathrm{au}$, the steady-state active-JFC count from \citet[][Figure~14]{Nesvorny2017}. Our adopted limit of $H_V < 29.11$ gives a total implied JFC population of $N(<29.11) \simeq4.76\times10^7$ objects.

The orbit sampling follows the same procedure as for Centaurs, but uses JFC orbit templates. We obtain the JFC templates by filtering the \citet{Nesvorny2019} Centaur steady-state catalog for JFC-range particles, following the \citet{LevisonDuncan1994} dynamical definition: Tisserand parameter with respect to Jupiter $2 < T_J < 3$ and perihelion distance $q < 2.5\,\mathrm{au}$, where
\begin{equation}
    T_J = \frac{a_J}{a} + 2\cos i\,\sqrt{\frac{a}{a_J}(1 - e^2)},
    \label{eq:tisserand}
\end{equation}
and $a_J = 5.2026\,\mathrm{au}$. We obtain $21{,}689$ JFCs from the original catalog of $26.1\times10^6$ objects ($0.083\%$ of the total). 

We assign each object a gray/blue or red spectral class with fractions $90\%$ and $10\%$, respectively. The strongly gray-dominated color distribution of JFCs reflects the thermal destruction of ultra-red surface organics during repeated perihelion passages \citep{Jewitt2015, WongBrown2016}. Gray objects are assigned the LSST broadband colors of the Bienor template and red objects the Pholus template, both from Table~2 of \citet{Murtagh2025}, the same templates used for Centaurs. We then convert to $H_r = H_V - (V-r)$ using $V-r = 0.35$ for blue objects and $V-r = 0.55$ for red objects. Each JFC is assigned a linear phase function with a fixed coefficient $\beta = 0.046\,\mathrm{mag\,deg^{-1}}$ applied uniformly across all filters, consistent with the mean phase function slope derived from JFC nucleus lightcurves by \citet{Kokotanekova2017}.

A random subset of $N_{\rm samp} = 10^7$ synthetic JFCs serves as a Monte Carlo sample of the full calibrated population ($N_{\rm tot} \simeq 4.76\times10^7$, $D > 10\,\mathrm{m}$), and all subsequent number counts are scaled by a factor of $4.76$.

\subsubsection{Scattering-TNOs}
\label{sec:scattering_population}
Our scattering-TNO model follows the TNO prescription of \citet{Kurlander2025} for the magnitude distribution, orbit templates, color templates, and phase-function treatment. We depart from it only by extending the magnitude distribution to our fainter size limit and generating an independent Monte Carlo subsample at that limit.

We draw each object's absolute magnitude $H_r$ following the OSSOS divot form of \citet{lawler2018ossos}:
\begin{equation}
    n(H_r) \propto
    \begin{cases}
        10^{\alpha_b H_r}, & H_r < H_b, \\[2pt]
        c^{-1}\,10^{\alpha_b H_b + \alpha_f (H_r - H_b)}, & H_r \ge H_b,
    \end{cases}
    \label{eq:scattering_divot}
\end{equation}
with bright- and faint-end slopes $\alpha_b = 0.9$ and $\alpha_f = 0.5$, break magnitude $H_b = 8.3$, and divot contrast $c = 3.2$. We normalize the distribution to $N(H_r < 8.66) = 90{,}000$, following \citet{lawler2018ossos}. Formally extrapolating the divot distribution to the $H_r \simeq 28.66$ limit gives a calibrated parent population of $N(H_r < 28.66) \simeq 6.23 \times 10^{14}$ objects.

We assign orbits from a template pool of $3128$ objects. The pool consists of a regrouped scattering subset of the CFEPS L7 model (CFEPS-L7; \citealt{Jones2006, Petit2011}) by \citet{Kurlander2025}, together with objects whose perihelia lie interior to Neptune, $q = a(1-e) < 30.06~\mathrm{au}$. For each synthetic object, we draw one template uniformly and copy its $(a, e, i)$, preserving the empirically motivated orbital structure of the scattering population. We randomize only the angular elements, drawing each independently from $\Omega, \omega, M \sim U(0^\circ, 360^\circ)$. This randomization reflects that scattering-TNOs are nonresonant and occupy no preferred orbital longitudes.

Each object is assigned either a blue or red TNO spectral template, with a blue fraction $f_{\rm blue} = 0.775$ \citep{Pike2023}, each based on the representative spectra of 2002~PN$_{34}$ and 1999~OX$_{3}$ respectively \citep{Seccull2018, DeMeo2010}. Conditioned on the assigned class, we draw the corresponding LSST colors from the TNO~blue and TNO~red entries of Table~3 in \citet{Kurlander2025}. We assign no phase function, as scattering-TNOs are observed over only $\sim0^\circ\text{--}3^\circ$ in phase angle, where phase effects are negligible.

We present the full $H_r < 28.66$ population with a Monte Carlo sample of $N_{\rm samp} = 2\times10^{7}$ and scale all resulting counts by a factor of ($2.975 \times 10^{7}$).

\subsection{Impactor identification workflow}

\paragraph{Overall workflow}
We identify impactors with a two-stage procedure. First, we screen the integrated source populations for planetary Hill-sphere close encounters. Second, we determine which of those encounters lead to impacts using two complementary methods: a periapsis-distance test, which gives a realized impact count, and a B-plane probability calculation, which gives an expected impact count for unbound encounters.

\paragraph{Motivation for the close-encounter layer}
The close-encounter layer is necessary because a binary impact flag is statistically inefficient and dynamically incomplete. Realized impacts are rare, so a direct impact count is dominated by small-number statistics and converges slowly. Hill-sphere close encounters provide a much larger sample from which impact probabilities can be estimated, including cases where no direct impact occurs in the finite integration.

This intermediate layer is also diagnostically useful, as it preserves the planet-relative state of each encounter, including encounter distance, relative velocity, and approach geometry. These quantities can reveal how planetary gravity shapes the encounter dynamics. A binary impactor flag collapses this information into a single number.

This layer also connects the simulation to operational impact monitoring. Monitoring systems track objects whose fitted orbits and uncertainties indicate potentially threatening close approaches \citep{Fuentes-Munoz_2023}. For Earth, the potentially hazardous asteroid criterion uses a minimum orbit intersection distance of $\mathrm{MOID}\leq0.05$~au \citep{1994hdtc}, well outside Earth's Hill radius of $\approx0.01$~au. A Hill-sphere crossing is therefore a closer approach than the threshold that flags a monitored object. The ratio of impacts to Hill-sphere encounters gives an empirical conversion efficiency that can be applied to the monitored population to estimate an upper limit on how many are expected to become actual impactors.

\paragraph{Implementation}

Our adopted implementation avoids two limitations of the more direct encounter-detection methods. A distance check embedded directly in the main \texttt{IAS15} integration through a \texttt{REBOUND} heartbeat callback can miss brief encounters, because the adaptive timestep is controlled by the global integration error. This is especially problematic for brief inner-planet encounters. Conversely, launching a separate high-resolution \texttt{IAS15} mini-integration for every Hill-sphere encounter resolves the sampling issue but is computationally expensive, requiring a new \texttt{REBOUND} instance, ephemeris reload, re-integration, and potentially thousands of sub-steps per event.

We therefore use the main integration to identify Hill-sphere encounters at a controlled, planet-specific cadence (Section~\ref{sec:input to encounters}), and then compute impact diagnostics analytically from the saved planet-relative state vectors (Section~\ref{sec:encounter to impactor}). This approach is computationally feasible for the full source populations and preserves the information needed for both periapsis-distance and B-plane impact estimates. We validate the encounter and impact classifications against refined \texttt{IAS15} mini-integrations and the Jet Propulsion Laboratory (JPL) Horizons ephemeris service \citep{giorgini2015Horison}. Details of this validation are given in Appendix~\ref{app:validation}.

\subsection{From Input Population to Close Encounters}
\label{sec:input to encounters}
We identify close encounters using a single geometric criterion. We define a close encounter of a planet as a test particle whose planet-centered distance falls within that planet's Hill radius ($R_{\mathrm{H},p}$). The Hill radius defines a spherical region inside which the planet's gravity dominates the Sun's, providing a natural boundary for recording close encounter states. For a planet of mass $m_p$ at semi-major axis $a_p$,
\begin{equation}
  R_{\mathrm{H},p} = a_p \left(\frac{m_p}{3M_\odot}\right)^{1/3},
  \label{eq:planet_hill_radius}
\end{equation}
and an object undergoes a close encounter with planet $p$ at time $t$ when
\begin{equation}
   \left\lVert \mathbf{r}_{\rm obj}(t) - \mathbf{r}_p(t) \right\rVert \le R_{\mathrm{H},p}.
  \label{eq:crosser_condition}
\end{equation}
The positions are taken from the 300-year N-body simulation of the input populations by \texttt{ASSIST} \citep{holman2023assist} and \texttt{REBOUND} \citep{Rein2012REBOUND}. The adopted Hill radius of each planet is listed in Table~\ref{tab:adopted_radii}. The duration of 300 years is chosen to be long enough to yield direct impacts while keeping the computational cost manageable. We reduce computational cost with a coarse and a fine screening procedure, detailed in Appendix~\ref{app:two stage}.

The Moon is treated separately because its Hill sphere lies entirely within Earth's, so we screen for lunar close encounters only among objects already flagged as Earth close encounters (see Appendix~\ref{app:moon}).

\begin{table}[htbp]
\centering
\caption{Adopted Hill-sphere and planet radii. The Hill radii set the close-encounter threshold; the planet radii set the impact threshold.}
\label{tab:adopted_radii}
\begin{tabular}{lccc}
\toprule
Target & $R_p$ (km) & $R_{\mathrm{H},p}$ (au) & $R_{\mathrm{H},p}$ ($10^6$ km) \\
\midrule
Mercury & 2440  & 0.0012 & 0.180 \\
Venus   & 6052  & 0.0067 & 1.002 \\
Earth   & 6378  & 0.0100 & 1.496 \\
Mars    & 3390  & 0.0070 & 1.047 \\
Jupiter & 71492 & 0.3550 & 53.1  \\
Saturn  & 60268 & 0.4120 & 61.6  \\
Uranus  & 25559 & 0.4700 & 70.3  \\
Neptune & 24764 & 0.7690 & 115.1 \\
\bottomrule
\end{tabular}
\end{table}

\subsection{From Close Encounters to Impactors}
\label{sec:encounter to impactor}
\subsubsection{Impactor Count from Periapsis Distance Calculation}
\label{sec:periapsis_method}
The most direct test of whether a close encounter is an impact is to compute the object's periapsis relative to the planet and compare it to the planet's physical radius.

For each close encounter we treat the object as following its osculating conic, the instantaneous Keplerian trajectory about the target planet which shares the same planet-centered position and velocity $(\mathbf{r}, \mathbf{v})$ as the full $N$-body trajectory at a given epoch. The inferred periapsis at the given epoch is
\begin{equation}
  r_{peri} = \frac{p}{1+e}, \qquad p = \frac{h^2}{\mu},
  \label{eq:periapsis_from_state}
\end{equation}
where $h$ is the specific angular momentum, and $e$ is the magnitude of the eccentricity vector. This is valid for both bound (elliptic) and unbound (parabolic and hyperbolic) cases. An encounter is a geometric impact candidate if the minimum periapsis during the Hill sphere entrance period is less than the threshold radius, i.e., $r_{peri,\min} = \min_t r_{peri}(t) \le R_{\rm th}$, where $R_{\rm th}$ is the solid-body radius for terrestrial targets and the adopted atmospheric reference radius for giant planets (1 bar). An object satisfying this condition is recorded as an impactor under the periapsis method. 



\subsubsection{B-Plane Impact Probability}
\label{sec:bplane_method}

\begin{figure}[ht!]
\centering
\includegraphics[width=0.95\linewidth]{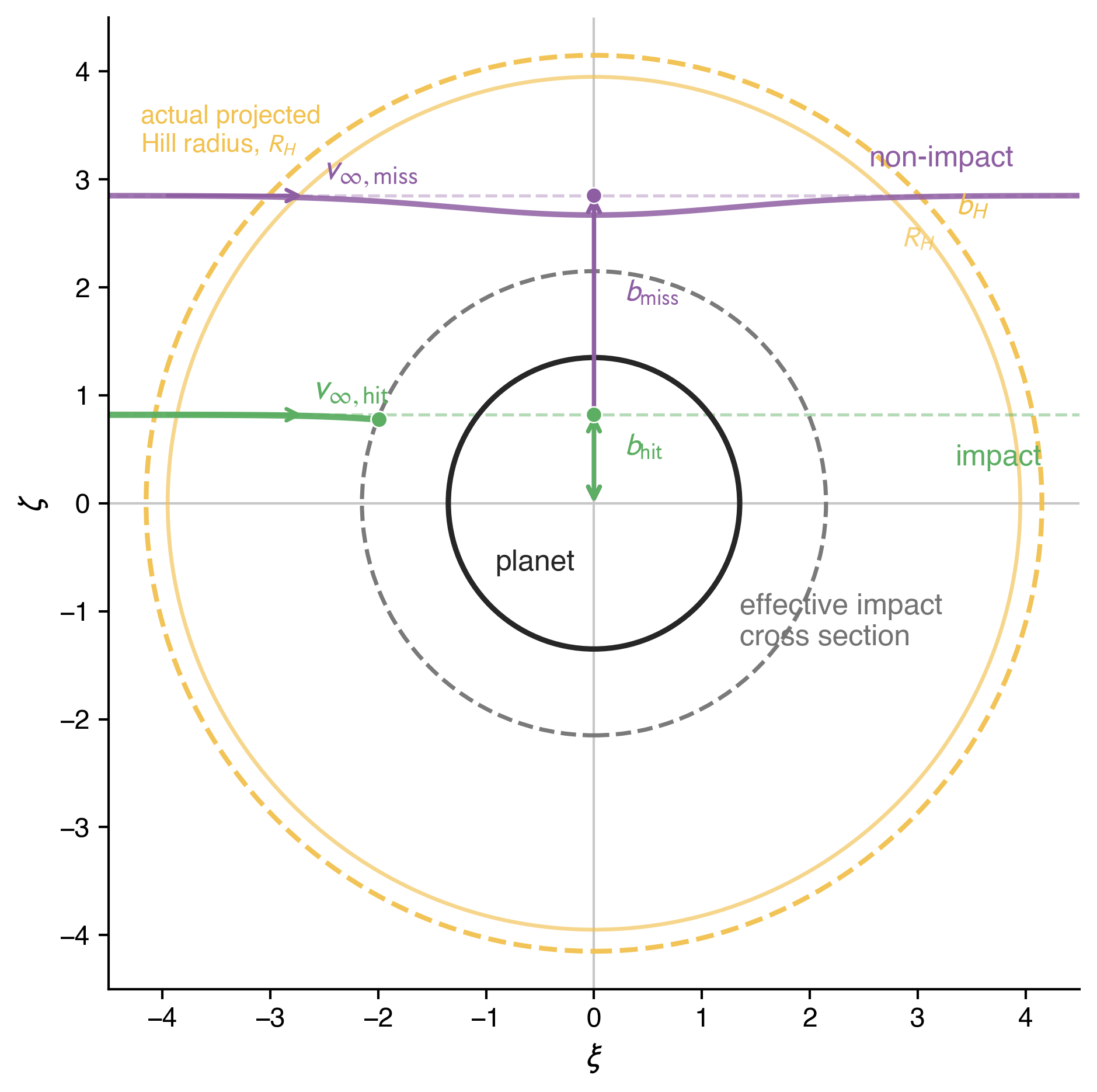}
\caption{Conceptual B-plane diagram illustrating the distinction between a close encounter and an impact. The incoming asymptotic trajectory is projected onto the encounter plane (B-plane), where the impact parameter $B$ determines whether the trajectory intersects the planet's effective collision cross section. Trajectories with sufficiently small B-plane offset result in impact, while larger offsets produce non-impacting close encounters.}
\label{fig:b_plane}
\end{figure}

The complementary B-plane method converts close-encounter samples into an expected number of impacts. The B-plane geometry is illustrated in Figure~\ref{fig:b_plane}. This yields a statistical estimate of the impact rate even when direct impacts are rare over the limited integration time.


For each close encounter within the Hill sphere, we measure its impact parameter on the B-plane \citep{farnocchia2019bplane}. This treatment assumes that the close approach is locally hyperbolic (unbound), with specific energy $\epsilon > 0$. Bound encounters are handled by the periapsis test of Section~\ref{sec:periapsis_method}.

The B-plane is the plane perpendicular to the object's incoming asymptotic velocity $v_\infty = \sqrt{2\epsilon}$ \citep{farnocchia2019bplane}. For a hyperbolic encounter with $v_\infty$, the corresponding impact parameter of an impact threshold $R_{\rm th}$ defines the gravitationally focused collision radius on the B-plane,
\begin{equation}
  B_{\rm imp} = R_{\rm th}\left[1 + \left(\frac{v_{\rm esc}}{v_\infty}\right)^2\right]^{1/2},
  \qquad v_{\rm esc} = \left(\frac{2\mu}{R_{\rm th}}\right)^{1/2},
  \label{eq:bimp}
\end{equation}
and the Hill radius $B_{\rm H} \simeq R_{\rm H}$ follows analogously (the focusing correction is negligible at $R_{\rm H} \gg R_{\rm th}$). Assuming close encounters are uniformly distributed over the Hill-crossing disk, the impact probability for one passage $k$ of object $i$ is the ratio of the two cross-sections,
\begin{equation}
P_{i,k} = \left(\frac{B_{{\rm imp},i,k}}{B_{{\rm H},i}}\right)^2
\simeq \left(\frac{R_{\rm th}}{R_{\rm H}}\right)^2\left[1 + \left(\frac{v_{\rm esc}}{v_{\infty,i,k}}\right)^2\right].
\label{eq:bplane_probability}
\end{equation}
The expected number of impacts from $N_{\rm cross}$ close encounters is $N_{\rm imp} = \sum_i P_i$, where one object's impact probability is $P_i = 1-\prod_k(1-P_{i,k})$ as one object may cross the Hill sphere more than once.


\section{Results}
\label{sec:results}
As the intrinsic impact rate is the focus of this paper, the results in this section refer to the calibrated, size-extended populations. The same analysis for the detection-limited, survey-facing populations is summarized in Appendix~\ref{app:original_population}.

\subsection{Comparison of Periapsis and B-plane Impact Estimates}
\label{sec:res_peri_bplane}
\input{master_v2_longrotatetable-paper}

In Table~\ref{tab:master_v2}, we compare the two impact-count estimators and assess which is more reliable under different circumstances. We report a realized object-level count by periapsis distance calculation, $N_{\rm imp,peri}$, and a statistical, population-level expectation of impactor number from objects with unbound orbits, $\langle N_{\rm imp}\rangle_{\rm B}$. The number of close encounters crossing a planet's Hill sphere, $N_{\rm cross}$, the number of unbound close encounters, $N_{\rm unb}$, and quantiles of periapsis distance and encounter speed are also provided as supplementary information. In this section we only briefly discuss qualitative uncertainties. Quantitative uncertainties are estimated and presented in Section~\ref{sec:uncertainties}.

Comparing $N_{\rm imp,peri}$ against $\langle N_{\rm imp}\rangle_{\rm B}$ shows that each estimator is reliable only under certain regimes. The two agree where encounters are unbound and unscaled, but diverge as the scaling factor and bound-encounter fraction grow. For unbound, unscaled populations such as NEOs, the two are broadly consistent, with Earth giving $5$ vs.\ $5.50$, Venus $6$ vs.\ $3.08$, and Jupiter $11$ vs. $7.14$ ($N_{\rm imp,peri}$ vs.\ $\langle N_{\rm imp}\rangle_{\rm B}$). The zero counts at Mercury and Mars are consistent with their small B-plane expectations ($0.08$ and $0.45$) over the 300-year integration. The JFC impactor estimations with a relatively small scaling factor of $\sim4$ likewise agree once uncertainties from counting statistics are taken into account. Jupiter gives $1636$ versus $1791.9$, and Saturn gives $33$ versus $16.3$. At Earth, the estimation based on periapsis distance is counting-limited, with a single raw impactor (scaled to $5$) versus a B-plane expectation of $0.46$. These comparisons support using the periapsis count as the primary realized impactor count and the B-plane estimate as a consistency check on the unbound-encounter ensemble.

As the scaling factor and bound object fraction increase, the two estimates diverge, and the B-plane count becomes the appropriate basis for interpreting impact rates. The divergence first appears in the MBA--Jupiter case. Even with a relatively small scaling factor ($25.65$), the periapsis method gives $N_{\rm imp,peri}=1334$, whereas the summed B-plane expectation is only $47.94$. The difference arises because only about one-fifth of the MBA--Jupiter close encounters are unbound ($N_{\rm cross}=2.73\times10^{5}$; $N_{\rm unb}=5.79\times10^{4}$), while the periapsis method also counts bound, low-energy, temporarily captured, or repeated-passage geometries inside Jupiter's Hill sphere. We discuss this Jupiter regime further in Section~\ref{sec:disc_jupiter_regime}. By contrast, Saturn's $0$ periapsis impacts versus a B-plane expectation of $2.140$ mainly reflects finite-sample and scaling uncertainty.

The Centaur population, with a scaling factor of $1195.04$, represents an intermediate case. The Centaur--Saturn channel has $N_{\rm imp,peri}=1195$ and $\langle N_{\rm imp}\rangle_{\rm B}=731.31$, with the full close-encounter ensemble classified as unbound. The Neptune and Uranus channels likewise have zero periapsis impacts versus B-plane expectations of $438.50$ and $621.36$. The periapsis count should therefore be treated as a noisy realization subject to large rare-event sampling uncertainty, while the B-plane estimate is used for the impact-rate calculation.

The scattering-TNOs carry the largest uncertainty, with a scaling factor of $2.975\times10^{7}$. This means the simulated subset represents only a minute fraction of the full population. Both the zero periapsis counts and the B-plane value are poorly constrained. The B-plane value places the slightly stricter constraint, and we therefore adopt it with its uncertainty. 

In summary, the periapsis count is the reliable measure for the inner, NEO-dominated planets, where encounters are mostly unbound and the scaling factor is unity. For the outer planets that have encounters from the scaled MBA, Centaur, JFCs, and scattering-TNO populations, the B-plane expectation should be adopted and the periapsis count only serves as a rare-event sampling check. 

\subsection{Uncertainties in the Two Impact Estimators}
\label{sec:uncertainties}
Because the two estimators sometimes disagree, we quantify their uncertainties to determine whether these differences are significant and what limits each method. Each impact estimator has two sources of uncertainty. The first is the Poisson counting uncertainty in the rare-event sample \citep{Gehrels1986}, referred to as counting uncertainty below. The same formalism provides an upper limit for cases where the number count is zero. The second is the uncertainty in the population scaling factor. Assuming these two contributions are independent, we combine them in quadrature to obtain the total uncertainty. Table~\ref{tab:unc_combined_full} lists the resulting uncertainties for every impact count in Table~\ref{tab:master_v2}. We find that where the direct and B-plane counts disagree, their $1\sigma$ or upper-limit ranges overlap, so the two estimators are not in conflict. The only exception is the MBA impactors at Jupiter, where the difference reflects the bound versus unbound impactor populations as discussed in Section~\ref{sec:res_peri_bplane}.


We illustrate the contribution of the scaling-factor uncertainty using two representative cases. The first is the NEO-Earth case, which has a small impact count and no population scaling ($f=1$). The second is the MBA--Jupiter case, which has a moderate raw count, requires population scaling, and shows a large difference between the two impactor estimators. We adopt a fractional scaling uncertainty of $\epsilon_f=0.2$ to illustrate its contribution to the combined uncertainty; we do not independently estimate $\epsilon_f$ in this work. 

For the NEO--Earth impactors, no scaling is applied ($f=1$). The periapsis method records $N_{\rm raw}=5$ impactors, giving a rate of $R=1.67\times10^{-2}~{\rm yr^{-1}}$. Its counting uncertainty is $1/\sqrt{5}=45\%$, which exceeds the assumed $\epsilon_f=20\%$, resulting in a combined uncertainty of $49\%$. The B-plane method gives a comparable expected count of $N_B=5.50$ and a rate of $R_B=1.83\times10^{-2}~{\rm yr^{-1}}$, with a counting uncertainty of $43\%$ and a combined uncertainty of $47\%$. The two estimates agree to within 10\% ($R_B/R\simeq1.1$), and both estimates are limited by rare-event counting statistics.

For the MBA--Jupiter impactors, a scale factor of $f=25.65$ is applied. The periapsis method records $N_{\rm raw}=52$ impactors (scaled to $1334$), giving a rate of $R=4.45~{\rm yr^{-1}}$. Its counting uncertainty is $1/\sqrt{52}=14\%$, below the assumed $\epsilon_f=20\%$, yielding a total uncertainty of $24\%$ ($R=4.45\pm1.08~{\rm yr^{-1}}$). The B-plane method gives a much smaller expected count of $N_B=1.87$ (scaled to $47.9$) and a rate of $R_B=0.16~{\rm yr^{-1}}$, with a counting uncertainty of $73\%$ and a combined uncertainty of $76\%$ ($R_B=0.16\pm0.12~{\rm yr^{-1}}$). The two estimators differ by a factor of $\sim28$, far exceeding either uncertainty. This gap reflects bound and multi-passage impactors that are counted by the periapsis method but absent from the unbound B-plane ensemble.

\input{uncertainty}

\subsection{Source-to-target Mapping}
\label{sec:res_source_target}
\begin{figure*}[t]
    \centering
    \centerline{\textbf{(a)} Hill sphere close encounters}
    \includegraphics[width=0.98\textwidth]{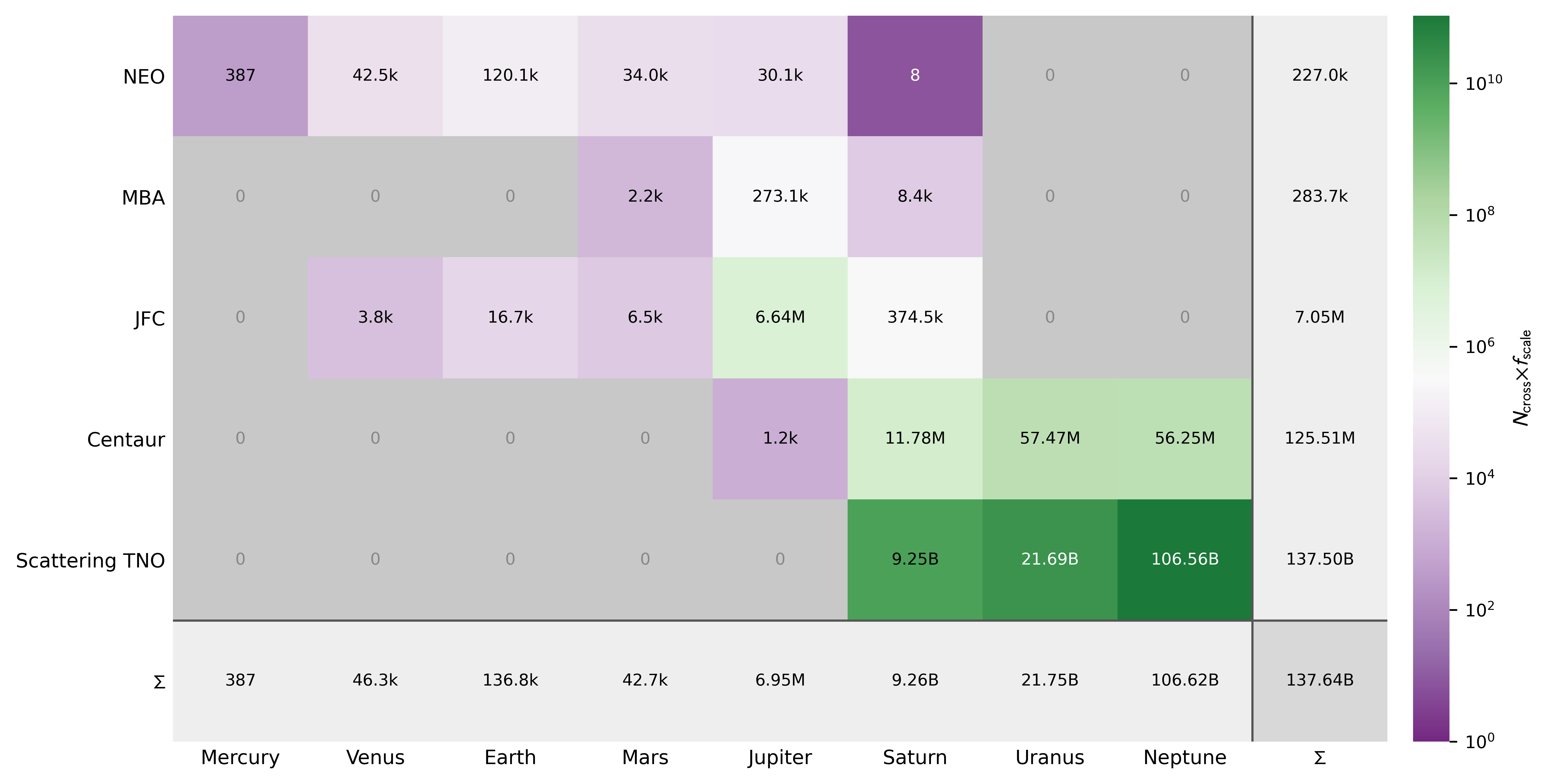}
    \vspace{-0.5em}

    \vspace{1.0em}
    \centerline{\textbf{(b)} Impactors}

    \includegraphics[width=0.98\textwidth]{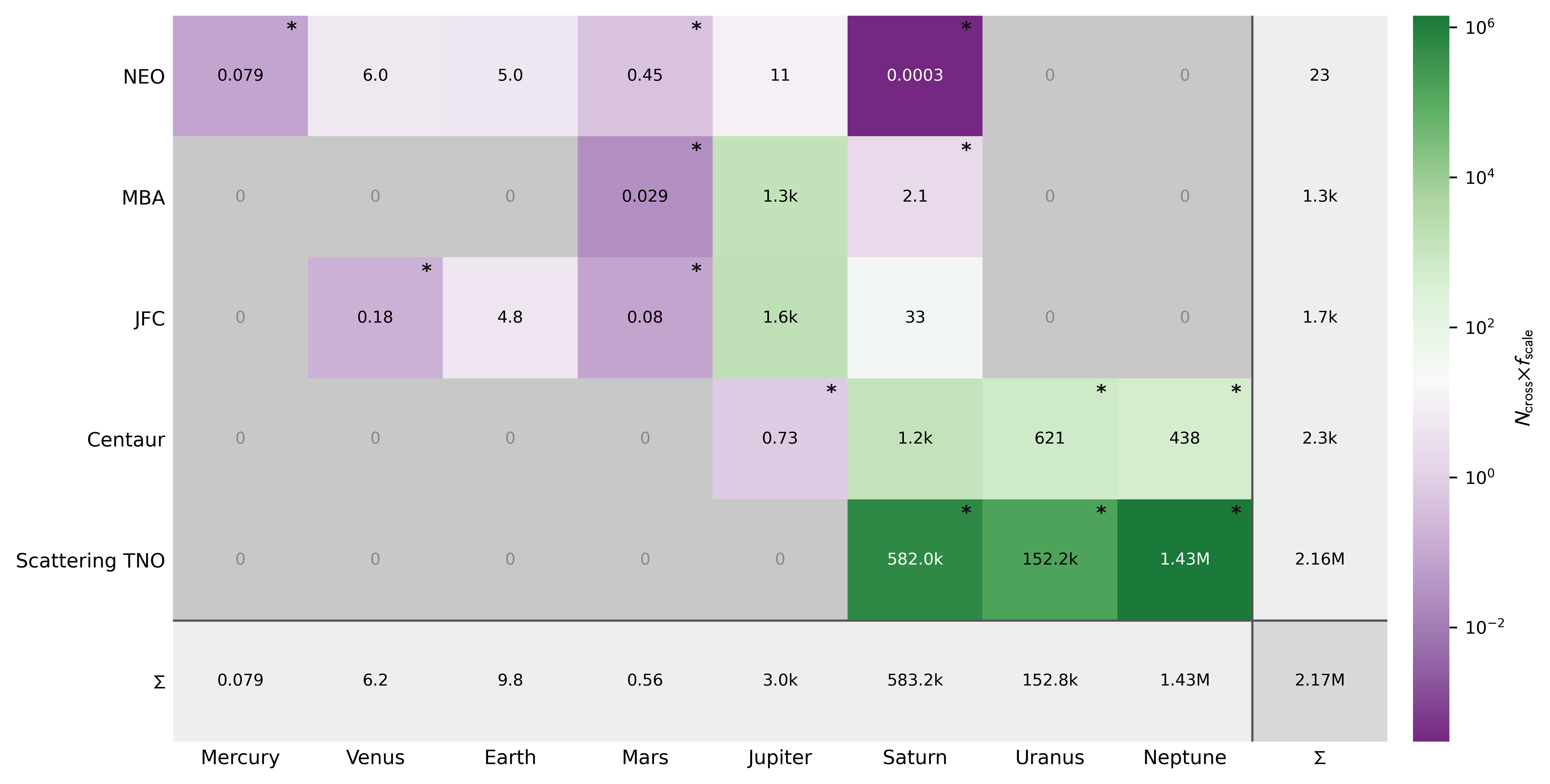}
    \vspace{-0.5em}

    \caption{
    Source--target conversion for the size-extended, calibrated populations.
    Panel (a) shows the number of Hill-sphere close encounters, $N_{\rm cross} f_{\rm scale}$.
    Panel (b) shows the number of impactors, $N_{\rm impact} f_{\rm scale}$.
    The close-encounter map shows broad source--target connectivity, while the impactor map isolates the smaller set of pathways that reach the physical collision regime.
    Asterisks indicate entries where no periapsis-confirmed impactors occurred in the finite simulation. For these entries, the plotted value is replaced by the corresponding B-plane collision expectation.
    }
    \label{fig:source_target_counts_ext}
\end{figure*}

\begin{figure*}[h]
    \centering

    \includegraphics[width=0.98\textwidth]{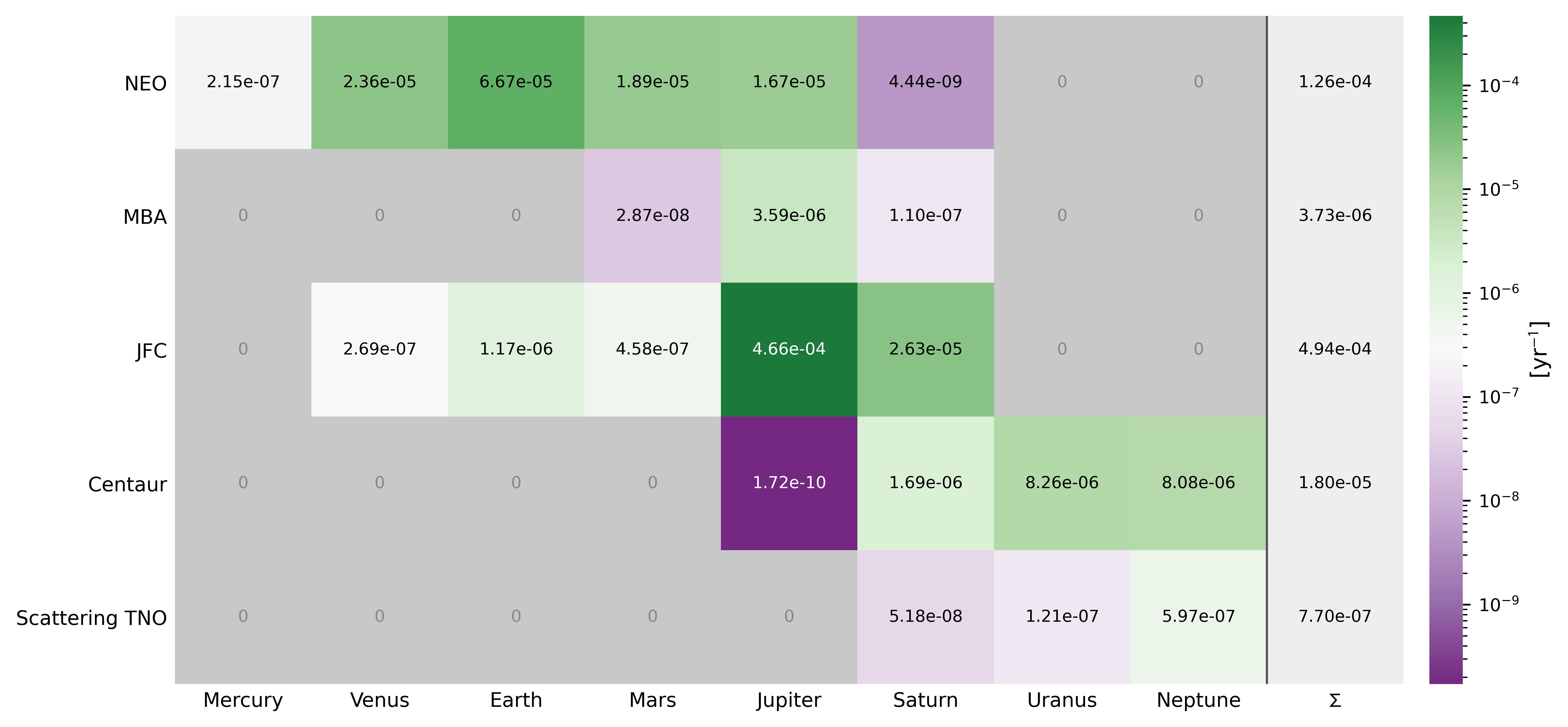}
    \vspace{-0.5em}
    \centerline{\textbf{(a)} Hill-sphere crossing rate}

    \vspace{1.0em}

    \includegraphics[width=0.98\textwidth]{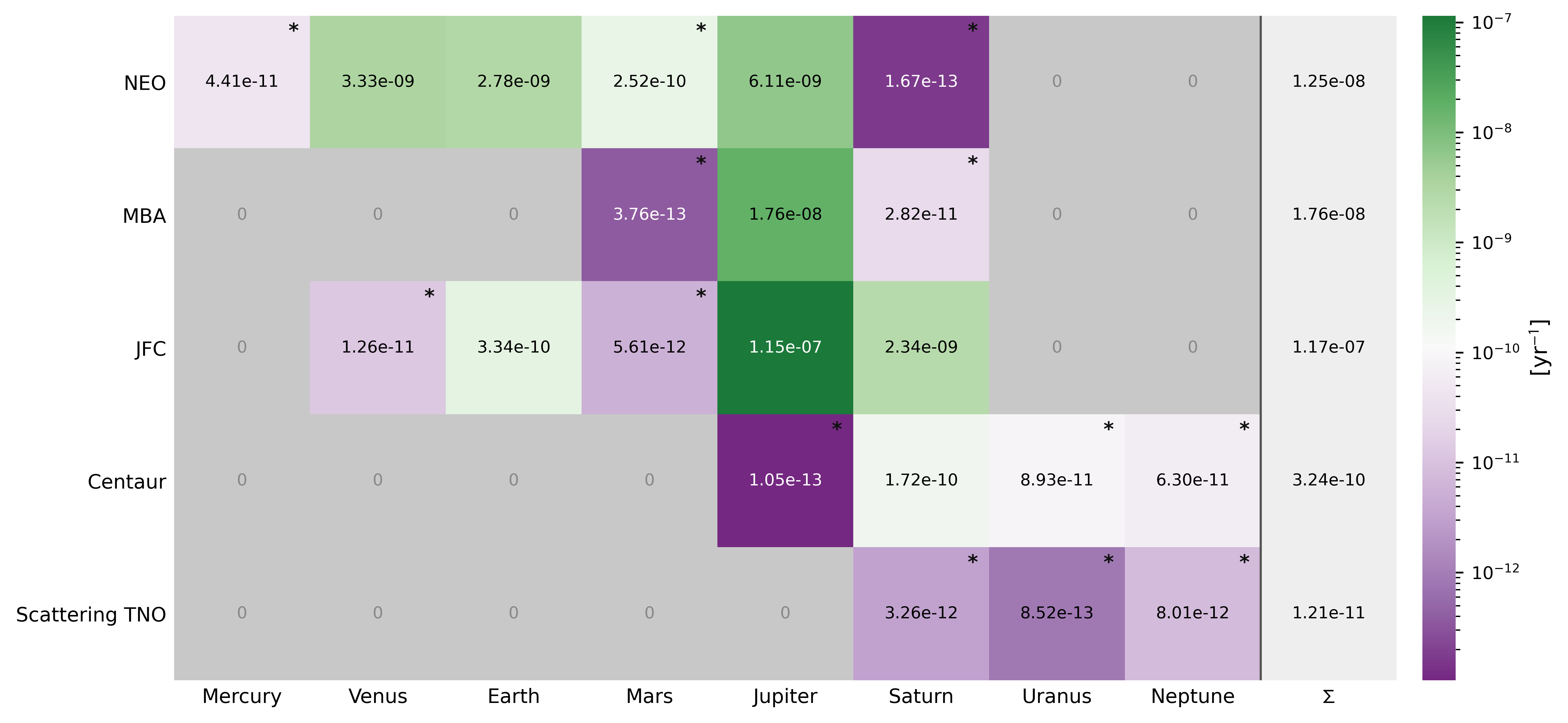}
    \vspace{-0.5em}
    \centerline{\textbf{(b)} Impact rate}

    \caption{
    Per-object dynamical efficiencies for the size-extended source populations.
    Panel (a) shows the annual Hill-sphere crossing rate per simulated object, $N_{\rm cross}/(N_{\rm sim}T)$.
    Panel (b) shows the annual impact rate per simulated object, $N_{\rm impact}/(N_{\rm sim}T)$.
    This normalization separates the dynamical efficiency of each source--target pathway from the adopted population scaling.
    Asterisks indicate entries where no periapsis-confirmed impactors occurred in the finite simulation. For these entries, the plotted value is replaced by the corresponding B-plane collision expectation.
    }
    \label{fig:source_target_rates_ext}
\end{figure*}

In Figure~\ref{fig:source_target_counts_ext} and Figure~\ref{fig:source_target_rates_ext}, we examine which planet receives the most close encounters and which source populations interact most efficiently with each target. We further investigate whether a large encounter or impact count reflects a large input population or a dynamically favorable source--planet pairing. Figure~\ref{fig:source_target_counts_ext} reports total counts, and Figure~\ref{fig:source_target_rates_ext} reports the per-object rates that isolate dynamical efficiency.

Figure~\ref{fig:source_target_counts_ext} shows that the total counts track the abundance of the input populations. The per-planet close-encounter counts increase from the inner to the outer Solar System, with Mars as the main exception. The counts rise from 491 at Mercury, to $4.6\times10^{4} -6.9\times10^{6}$ from Venus through Jupiter, and to $8.88\times10^{9}-1.0757\times10^{11}$ from Saturn through Neptune. These values carry population scaling from every source except the NEOs, so the largest counts also carry the largest uncertainty, most severely at Uranus and Neptune where scattering-TNOs dominate the input. Impact counts follow a similar trend. Combining the periapsis counts with the B-plane expectations as discussed in Section~\ref{sec:res_peri_bplane}, Neptune records the largest expected impact count, of order $10^{6}$, and the largest uncertainty (set by the 456\% scattering-TNO uncertainty). Saturn follows at $\sim5.8\times10^{5}$, Uranus at $\sim1.5\times10^{5}$, and Jupiter at $\sim3.0\times10^{3}$. Mercury and Mars fall below one expected impact, at $0.08$ and $0.56$ respectively. Source abundance explains the broad inner-to-outer increase in both visitors and impactors, as the most abundant populations, scattering-TNOs and Centaurs, interact with the outer Solar System planets most.

 


Figure~\ref{fig:source_target_rates_ext} shows the same source--target pairs in terms of per-object dynamical efficiency. This complementary plot normalizes the close-encounter and impact counts by the adopted population size, including the scaling factors, and by the total integration time. Per-object close-encounter rates peak for JFCs at Jupiter, followed by NEOs at Earth, Venus, Jupiter, and Mars, and by Centaurs at Uranus and Neptune. This ordering matches the known dynamical activity of the JFC and NEO populations. Scattering-TNOs dominate the outer-planet totals but have lower per-object rates than Centaurs, most clearly at Neptune. This suggests that individual scattering-TNOs have lower encounter efficiencies than NEOs, JFCs or Centaurs. Together, input-population abundance sets the overall scale of planetary interactions, while dynamical efficiency sets the impactor supply at Jupiter and Earth.


\begin{figure*}[h]
    \centering
    \includegraphics[width=0.92\textwidth]{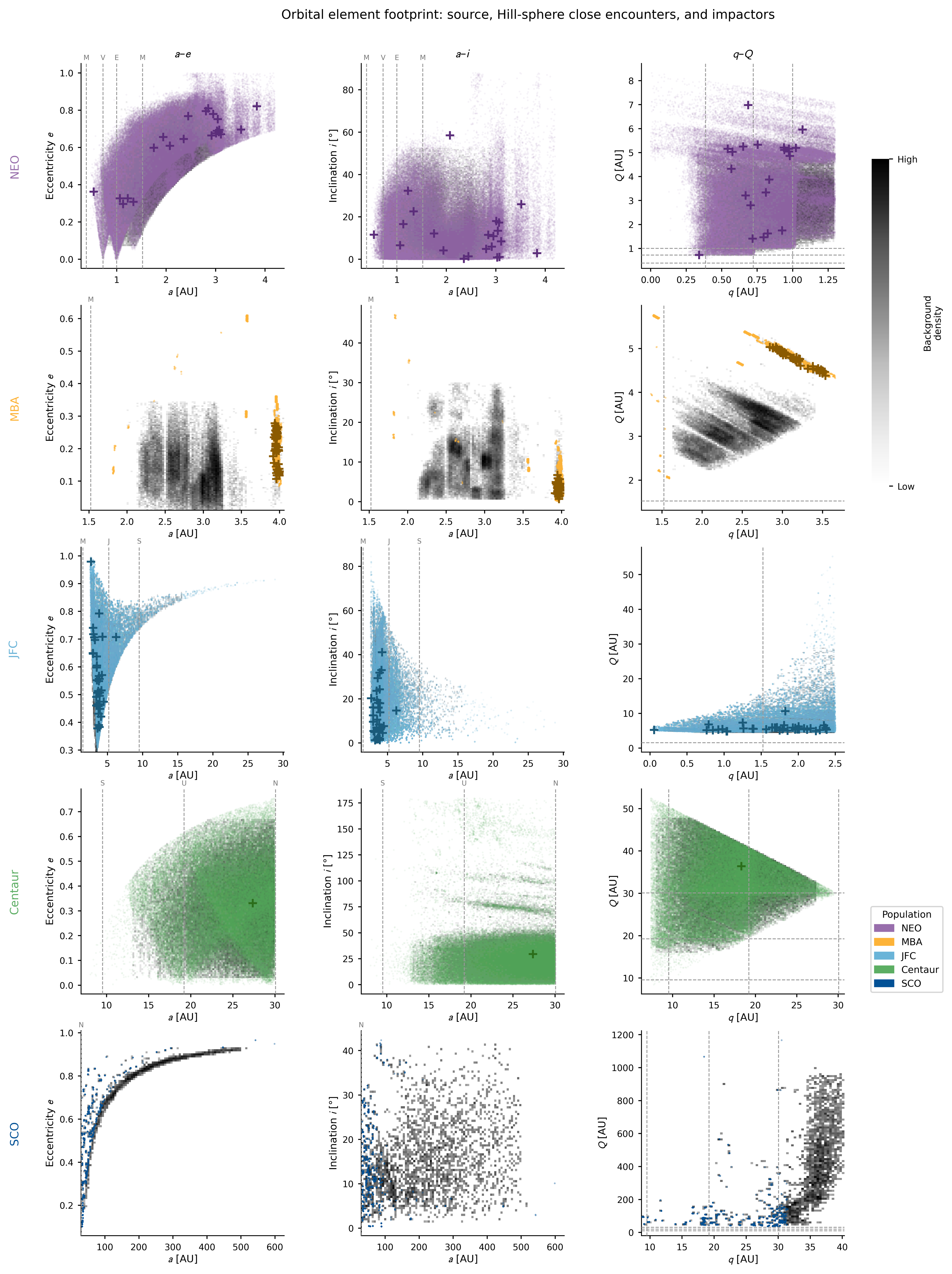}
    \caption{Orbital-element footprint of the size-extended Hill-sphere close encounters. Rows correspond to source populations and columns show the $a$--$e$, $a$--$i$, and $q$--$Q$ projections. The background density shows the parent source population, colored points show Hill-sphere close encounters, and plus symbols mark periapsis-confirmed impactors. The figure illustrates the orbital filtering imposed by planetary encounters: NEO impactors occupy the terrestrial-planet-reaching subset of the NEO distribution, MBA impactors are concentrated in the Jupiter-coupled outer-belt tail, JFC impactors are drawn from the low-perihelion, high-eccentricity region, and outer-Solar-System close encounters are selected from low-perihelion Centaur and scattering-TNO orbits.}
    \label{fig:orbital_footprint_ext}
\end{figure*}

\begin{figure*}[tp]
    \centering
    \includegraphics[width=1\textwidth]{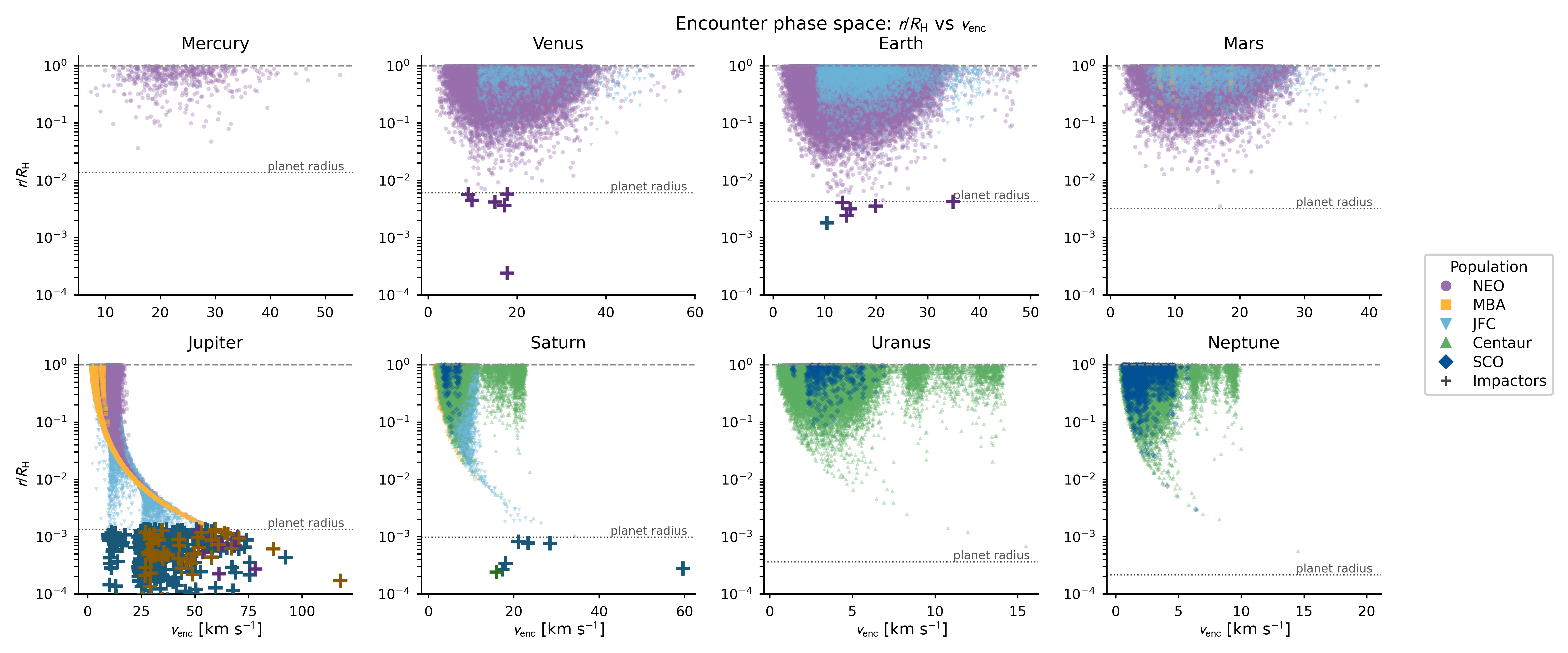}
    \caption{Close-approach depth and encounter speed for Hill-sphere close encounters. Each panel shows one target planet. The y-axis is closest approach normalized by the Hill radius, \(r/R_{\rm H}\), and the x-axis is planet-centered encounter speed, \(v_{\rm enc}\). The dashed line marks the Hill radius and the dotted line marks the planet radius. Points are colored by source population, and plus symbols mark periapsis-confirmed impactors.}
    \label{fig:encounter_phase_ext}
\end{figure*}

\subsection{Orbital-element Distributions of Input Populations, Close Encounters, and Impactors}
\label{sec:res_orbital_filtering}
Figure~\ref{fig:orbital_footprint_ext} plots the orbital-element distributions of close encounters and impactors of all five source populations, highlighting the dynamically active regions where close approaches are most likely to lead to collisions. Each panel shows these distributions relative to their source population, with dashed reference lines marking the mean orbital elements of each target planet. The NEO close-encounter sample spans much of the parent distribution, reflecting that many NEO orbits already cross or approach the terrestrial-planet region. The periapsis-confirmed NEO impactors are drawn from this same planet-reaching phase space. Terrestrial-planet impactors come from orbits whose perihelia and aphelia overlap the inner planets, whereas the Jupiter impactors come from the high-eccentricity, large-aphelion tail. The MBA close-encounter sample is more localized, with most of them drawn from the low- to moderate-inclination outer-belt tail near the Jupiter-crossing boundary, producing recurring encounters. A smaller high-eccentricity, low-perihelion component also reaches Mars. The JFC close encounters trace the low-perihelion ($q \lesssim 2.5$~au) region of the JFC distribution, whose perihelia reach the terrestrial-planet region while their aphelia remain near Jupiter. Most Centaur close encounters span the perihelion-to-aphelion ranges of Uranus and Neptune, and the scattering-TNO encounters reach inward to the $q<30$~au Neptune boundary. We discuss the cross-Solar-System orbital structure grouped by planets, along with its implications, in Section~\ref{sec:disc_across_solar}.

\subsection{Close-approach Depth, Velocity, and Impact Speed}
\label{sec:res_encounter_geometry}

To understand the dynamics near an impact event, Figure~\ref{fig:encounter_phase_ext} plots closest-approach distance against relative speed for close encounters and impactors, with distance normalized by the Hill radius. Most Hill-sphere close encounters remain shallow, at $r/R_{\rm H}$ near unity, and the periapsis-confirmed impactors occupy the deep-encounter tail as expected. This tail is clearest at Venus and Earth where the NEO impactors lie at $r/R_{\rm H}\lesssim10^{-1}$, and at Jupiter where the NEO, MBA, and JFC impactors reach the smallest normalized depths. Across the well-sampled panels the fastest encounters never reach the smallest $r/R_{\rm H}$, whereas the slowest do, the expected signature of gravitational focusing.

The Jupiter panel is dynamically distinct, as expected based on previous discussions in Section~\ref{sec:res_peri_bplane}. The MBA--Jupiter close encounters form a low-speed, deeply penetrating sequence, and the impactors lie along the deepest part of this sequence. By contrast, the NEO-- and JFC--Jupiter close encounters cluster off the focusing curve, and their impactors include higher-speed encounters. Jupiter's distinct dynamics are discussed in Section~\ref{sec:disc_jupiter_regime}. Saturn shows Centaur and JFC impactors at small $r/R_{\rm H}$ and moderate encounter speed, while Uranus and Neptune show large but shallow close-encounter populations, consistent with the absence of periapsis-confirmed impactors. Thus, for the outer planets, large Hill-sphere close encounter counts do not automatically imply sampled physical collisions over the finite integration baseline. 

\begin{figure*}[tp]
    \centering

    \includegraphics[width=1\textwidth]{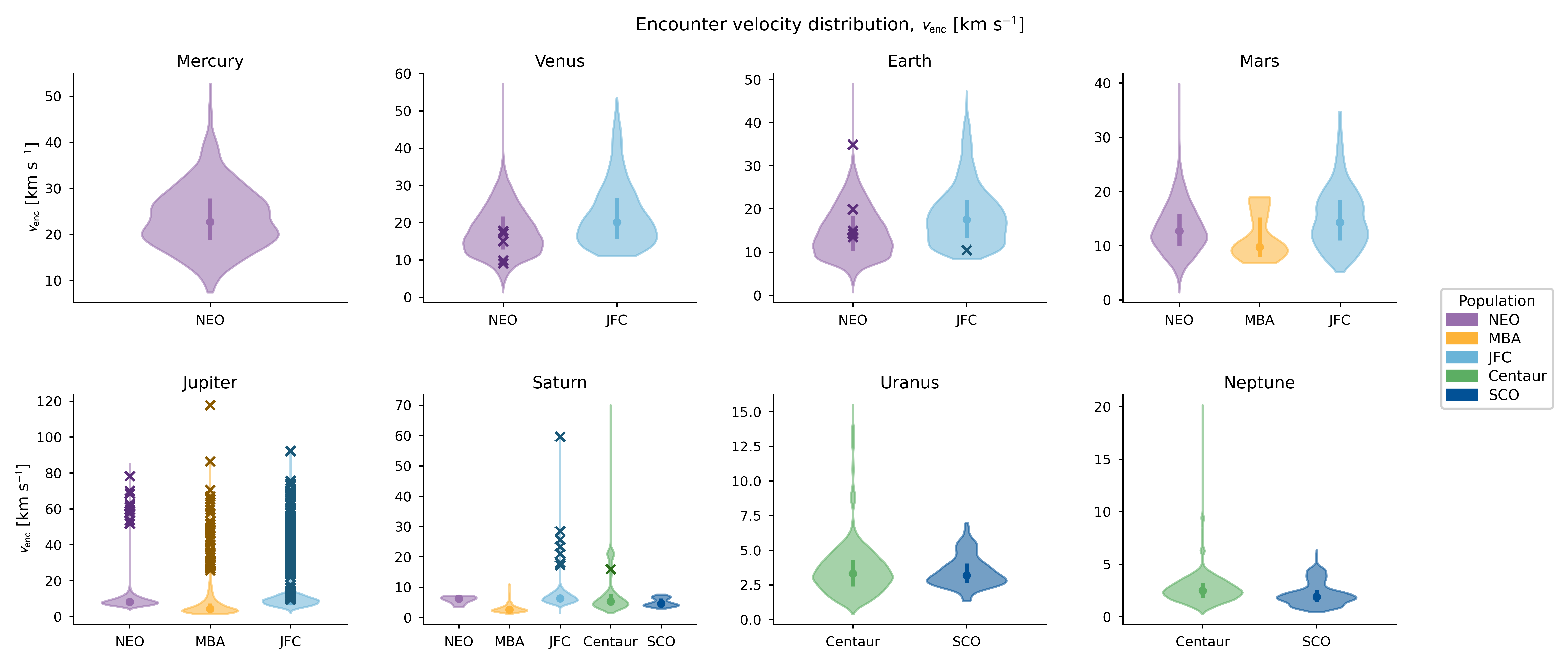}

    \caption{Speed distributions for the Hill-sphere close encounters and impactors by source and target. For non-impacting Hill-sphere encounters, the speed is the instantaneous relative speed $v_{\rm enc}$ evaluated at the sampled timestep of closest approach. For flagged impactors, the speed is the two-body periapsis speed $v_{\rm peri}=\sqrt{GM\,(2/r_{peri} - 1/|a|)}$ from the osculating orbit fit at $r_{peri}$.}
    \label{fig:encounter_depth_velocity_ext}
\end{figure*}

The same trends appear in the relative-velocity distributions in Figure~\ref{fig:encounter_depth_velocity_ext} and in the encounter-speed columns in Table~\ref{tab:master_v2}. Encounter speeds at the terrestrial planets reach tens of kilometers per second for the NEO and JFC populations. At Uranus and Neptune, the Centaur and TNO encounters are far slower, only a few kilometers per second, with Saturn intermediate. Mean impactor speeds span from $\sim 20\,\mathrm{km\,s^{-1}}$ for Venus and Earth impacts, to high-speed tails extending beyond 60 $\mathrm{km\,s^{-1}}$ for Jupiter. Within the Jupiter sample, bound MBA impactors occupy a much wider velocity range and extend to lower speeds than the unbound impactors from NEOs and JFCs. These velocity differences provide a basis to convert observables such as energy or flux to impact diameters in Paper II. 

\subsection{Temporal Evolution and Statistical Convergence}
\label{sec:res_time_evolution}

\begin{figure*}[tp]
    \centering
    \includegraphics[width=0.99\textwidth]{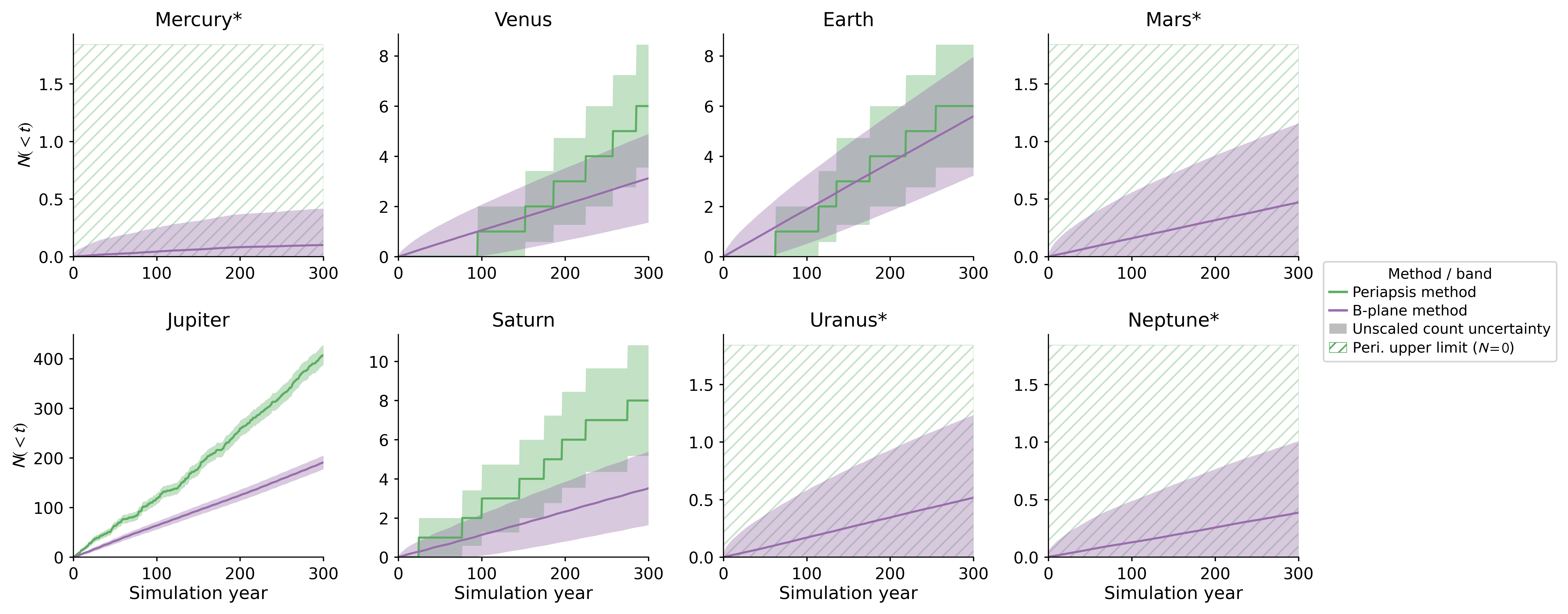}
\caption{Cumulative number of simulated impacts $N(<t)$ as a function of simulation year, shown in raw (unscaled) counts for all eight planets. Green curves and shaded bands show the periapsis-method estimate with its $1\sigma$ Poisson counting uncertainty. Purple curves show the B-plane expected count $\Sigma f_B(<t)$ with its Poisson-binomial counting uncertainty. Uncertainty derivation can be found in Section~\ref{sec:uncertainties}. An asterisk in the panel title marks these null-detection planets.}
    \label{fig:time_evolution_ext}
\end{figure*}

In Figure~\ref{fig:time_evolution_ext}, we track the cumulative number of Hill-sphere close encounters and impactors over 300 years of integration to test whether the counts are stable enough to interpret as rates. The close-encounter curves are nearly linear over the 300-year integration, which justifies the use of time-averaged close-encounter rates for the high-count channels. The impactor curves are more step-like due to the rare impact events. Impacts at Venus and Earth appear as a few discrete events, whereas Jupiter impacts accumulate the most steadily, making it a well-defined impact rate. For planets with zero periapsis-detected impactors, which are Mercury, Mars, Uranus, Neptune, the B-plane estimate rises steadily but stays below 1 by the end of the integration. Its slope nevertheless provides an estimate of the impact rate. 

\section{Discussion}
\label{sec:discussion}

\subsection{Close Encounters and Impactors across the Solar System}
\label{sec:disc_across_solar}

\begin{figure*}[tp]
\centering
\begin{subfigure}{1\textwidth}
    \centering
    \includegraphics[width=\textwidth,height=0.21\textheight,keepaspectratio]{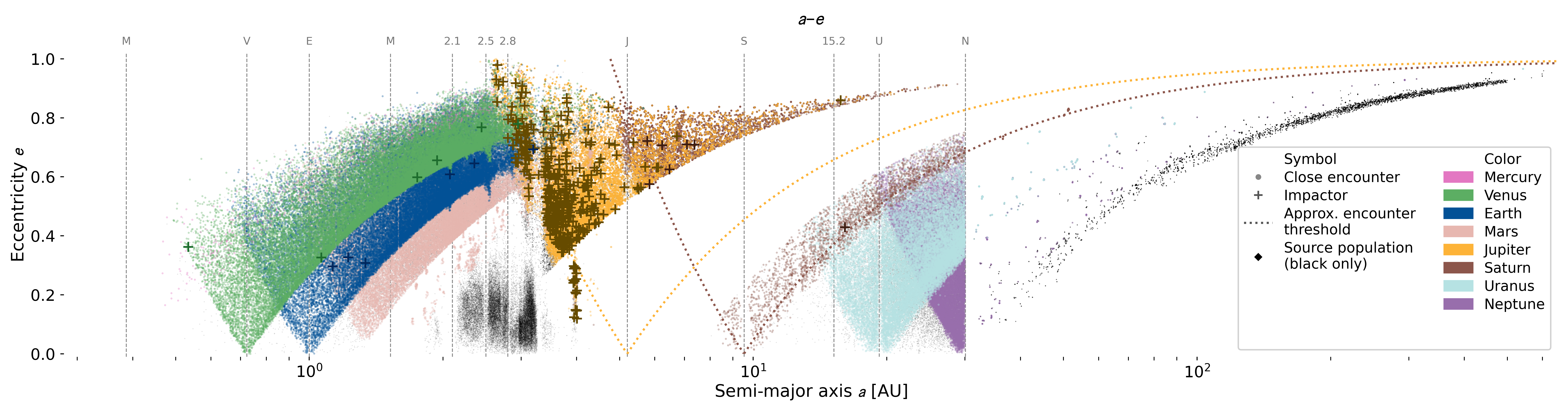}
    \caption{}
    \label{fig:orbital_ae_4}
\end{subfigure}
\vspace{0.6em}
\begin{subfigure}{1\textwidth}
    \centering
    \includegraphics[width=\textwidth,height=0.21\textheight,keepaspectratio]{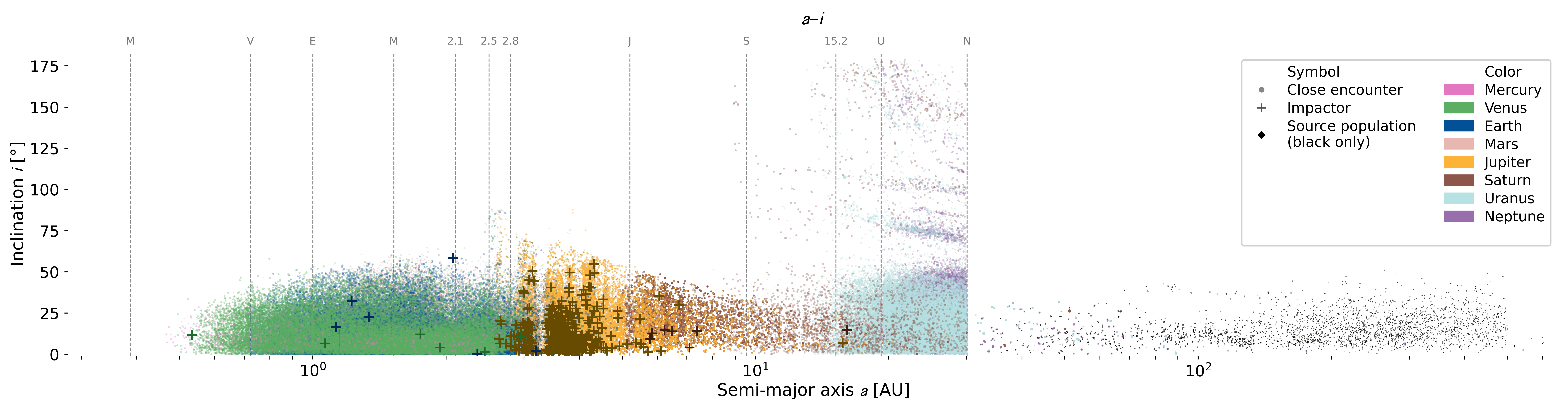}
    \caption{}
    \label{fig:orbital_ai_4}
\end{subfigure}
\caption{
Orbital footprints of the combined simulated populations, close encounters, and realized impactors across the Solar System, with objects color-coded by the target planet encountered.
Panel (a) show semi-major axis--eccentricity space and panel (b) show semi-major axis--inclination space.
In each panel, the faint gray points show the full input populations for reference only.
Colored points denote close-encounter objects, and crosses denote realized impactors.
Vertical dashed lines mark the eight planetary semi-major axes and resonances. 
Objects that encounter multiple planets are plotted multiple times, once for each target planet. 
}
\label{fig:orbital_footprints_combined}
\end{figure*}

The orbital-element distributions by target planet reveal both geometric and dynamical patterns in the close-encounter and impactor populations. Geometrically, the clearest structure appears as V-shapes in (a)--(e) space when objects are grouped by the planet they encounter, as shown in Figure~\ref{fig:orbital_footprints_combined}. An object can encounter an exterior planet if its aphelion reaches the planet's orbit,
\[
Q = a(1+e) \gtrsim a_p,
\]
or an interior planet if its perihelion reaches the planet's orbit,
\[
q = a(1-e) \lesssim a_p,
\]
where $a_p$ is the planet's semi-major axis. These two limits define the two branches of the V-shaped envelope which traces the geometry required for an orbit to approach a given planet.

This geometric view explains how different source populations contribute to close encounters. Dynamically active populations, such as NEOs and scattering-TNOs, have orbital distributions that already overlap much of the V-shaped region. More stable populations overlap only in restricted zones, for example, for MBAs, the outer belt and high-eccentricity inner-belt regions. We use dashed curves to trace the incomplete V-shapes.

The density of close encountering objects within V-shaped envelopes provides information about the dynamical routes of close encounters. In Figure~\ref{fig:orbital_ae_4}, the vertical spikes near $\simeq 2.1, 2.5,$ and $2.8$ au appear in Venus, Earth, and Mars encounters. The spikes suggest perturbations in eccentricity, shown as the relatively dense regions near the same three semi-major-axis values in Figure~\ref{fig:orbital_footprint_ext}. These regions are consistent with known main-belt escape routes. The 2.1 au subgroup coincides with $\nu_6$ secular resonance \citep{morbidelli_nesvorny1999, bottke2002}, an inner main-belt escape region. The other two subgroups around 2.5 au and 2.8 au are consistent with the 3:1J and 5:2J resonances about Jupiter, which are the classical middle main-belt escape channels \citep{wisdom1983, gladman1997}. The same regions appear as ``Kirkwood gaps" in the MBA source distribution, suggesting that these resonances act as efficient delivery channels from the main belt into planet-crossing orbits. The clustering of realized impactors near these regions, especially for the ones from MBAs, further supports resonant delivery as an important pathway \citep{bottke2002, granvik2018}. Similar structure appears in the outer Solar System. In particular, the gap near $a \simeq 15.2$ au, associated with the Saturn 2:1 resonance \citep{tiscareno_malhotra2003}, is visible in the outer-planet encounter populations. These patterns show that close encounters are shaped by simple orbit-crossing geometry, and the density of close encounters and impactors within that geometry is controlled mostly by resonances.

The overlap between the V-shaped regions marks another interesting group of objects that are most likely perturbed by more than one planet. Such multi-planet encounters are rare in our simulations. Only $1.65\%$ of close-encounter objects pass near two planets, and the most common pair is Earth--Venus (4,783 objects). Only seven objects encounter three planets, all in the Earth--Mars--Venus group. Among the realized impactors, there is one that crosses multiple planets: an MBA-originated Jupiter impactor that also encounters Saturn, with $a=3.963$ au, $e=0.207$, and $i=3.15^\circ$.

\subsection{Jupiter as a Dynamically Distinct Case}
\label{sec:disc_jupiter_regime}
Jupiter is the most dynamically distinctive target among all the planets. Its strong gravity drives a high impact rate, while its large Hill sphere, high escape speed, and strong gravitational focusing allow it to trap slow planet-centered trajectories that loop repeatedly before they either escape or strike \citep{suetsugu2013}. Jupiter therefore has both fast hyperbolic impacts from a passing flux and multi-passage impacts driven by temporary capture. This two-component picture is supported by the disagreement between the two impactor-count estimators with uncertainties (Section~\ref{sec:res_peri_bplane} and Section~\ref{sec:uncertainties}), and the speed and encounter depth distribution (Section~\ref{sec:res_encounter_geometry}). The coexistence of fast unbound impactors and slow temporarily captured ones is not seen for the other planets in our results. 

This behavior has clear precedents at Jupiter. Shoemaker--Levy~9 showed that Jupiter can tidally disrupt and then accrete a temporarily captured object, and quasi-Hilda comets provide a known pathway into prolonged temporary capture within Jupiter's Hill sphere \citep{Movshovitz_2012,Ohtsuka2008KushidaMuramatsu}. The impact flux is also high enough to measure. Optical flashes from $5-20$ m objects imply a selection-corrected rate of order tens per year \citep{hueso2018small}, and ultraviolet detections from the Juno mission suggest a much higher meter-class rate, though it is from a single event with large conversion uncertainty \citep{giles2021detection}. Jupiter is therefore a transition regime, where the impactor dynamics and impact-estimation assumptions both differ from those at other planets, and its observational opportunities are ample.

\subsection{Implications for the Earth source--target Mismatch}
\label{sec:disc_earth_mismatch}

The NEO--Earth channel provides a useful calibration point for comparing our intrinsic dynamical rates with the survey-population-derived decameter impact rate discussed by \citet{Chow2025decameter}. Over the 300 yr integration, we find 5 periapsis-confirmed $>10$ m impacts from the NEO population and a B-plane expectation of 5.498. These correspond to impact rates of $1.67\times10^{-2} \mathrm{yr^{-1}}$ and
$1.83\times10^{-2} \mathrm{yr^{-1}}$, or one NEO impact every $\sim$55--60 yr. As discussed in Sections~\ref{sec:res_peri_bplane} and \ref{sec:uncertainties}, the NEO--Earth impact rate is internally well resolved. The corresponding external comparison is the decameter Earth-impact rate inferred by the debiased telescopic NEO population from \citet{Chow2025decameter}, predicting decameter Earth impacts ($26.5 \le H \le 28$) roughly once every 20--40 yr ($R_{\oplus,\rm survey}\sim0.025$--$0.05\,\mathrm{yr^{-1}}$). Our rate is lower by a factor of about $1.4-3$, given that the size threshold, $H$--diameter conversion, and albedo assumptions are not exactly the same between the two calculations. Even so, this near-agreement makes the NEO--Earth channel a useful benchmark for the broader Solar System comparison, and the larger offsets seen at other planets may then signal a more general mismatch between source populations and the way each target is observed. 

The JFC population contributes one raw periapsis-confirmed Earth impact, which becomes $\sim5$ impacts after $f=4.756$ scaling. This brings the total periapsis-distance-derived impact count to 10, or $3.3\times10^{-2}\,\mathrm{yr^{-1}}$ (one impact every $\sim$30 yr). This cometary contribution is poorly constrained given its $100\%$ counting uncertainty, the absence of cometary-activity modeling, and a markedly lower B-plane expectation of 0.46. Existing observational evidence for a cometary contribution to Earth impactors is similarly ambiguous. Among the 14 USG impactors analyzed by \citet{Chow2025decameter}, none was identified as cometary. Six, however, have Tisserand parameters near the nominal asteroid--comet boundary ($T_J\simeq3$), and could be either JFC-like asteroids or dormant cometary objects. Therefore, our simulation suggests that a JFC-like component could increase the modeled Earth-impact rate, but neither its magnitude nor its origin is determined. Larger JFC simulations are required to improve the counting statistics, test the periapsis--B-plane discrepancy, and predict the orbital and compositional signatures for better comparison with the observed rate.

\subsection{Limitations and Other Uncertainties}
\label{sec:disc_limitations}

\subsubsection{Albedo Assumptions and Population Scaling}

One of the largest uncertainties in our pipeline is the conversion from absolute magnitude to physical diameter that depends on the assumed albedo. For fixed $H$, Equation~\ref{eq:h_to_d} gives $D\propto p_V^{-1/2}$ \citep[][]{PravecEtAl2012}. If the cumulative size distribution follows $N(>D)\propto D^{-\alpha}$, then rescaling the assumed albedo from $p_0$ to $p_1$ changes the inferred population above a fixed diameter threshold by
\begin{equation}
    \frac{N_1}{N_0}=\left(\frac{p_1}{p_0}\right)^{-\alpha/2}.
\end{equation}
For $\alpha=2$, shifting $p_V=0.14$ to $p_V=0.04$ increases the population size by a factor of $\simeq3.5$; for $\alpha=2.5$, the factor becomes $\simeq4.8$. Plausible albedo choices shift the final impact rates by factors of a few. 

The uncertainty in this scaling factor also dominates the total error budget. The estimated 3.5-fold difference corresponds to a $\sim$350\% uncertainty on the scaling factor, far larger than the counting uncertainty in most cases. It flattens the total uncertainty to $\sim$350\% for every well-sampled case, such as the Jupiter impactors (previously at $\sim20$\%) and the NEO impactors ($\sim$30--45\%). In these cases, running a larger sample of simulations no longer improves the results. Future work requires more precise albedo and size-frequency slope distributions for each source reservoir.

\subsubsection{Integration Limitations for Co-planar Objects}

A few highly co-planar asteroids pose an integration bottleneck in our MBA and JFC runs. When a low-inclination object undergoes a close, strongly perturbed encounter, the adaptive integrator collapses its timestep to resolve the trajectory, and the per-object cost rises sharply. This prevents us from integrating the full sample to convergence. We document the affected cases in Appendix~\ref{app:s100brmpa} and our reported impact rates are limited by the absence of highly co-planar objects.

\subsubsection{Other Unquantified Limitations}
Below are limitations whose effects we have not quantified. 
\begin{itemize}
    \item \textbf{Missing physical processes.} The present model omits non-gravitational forces that act over long timescales, such as Yarkovsky drift for small asteroids \citep{Vokrouhlicky2015} and outgassing for active comets \citep{Yeomans2004}. The JFC, Centaur, and Scattering-TNO populations are propagated as point masses under gravitational dynamics, whose cometary effects can cause notable differences on a timescale of 300 years. Our model also omits fragmentation, rotational breakup \citep{Walsh2018}, and tidal disruption \citep{Richardson1998}. These processes can alter the derived impact rates. Their importance depends on the size range, the target body, and the observational comparison. We defer this to future work in our companion paper.

    \item \textbf{Finite integration time.} The finite integration length is a practical limit set by computational cost. Appendix~\ref{app:simulation run time} gives detailed runtimes. The same limit constrains how well we resolve lunar encounters within the Earth--Moon system. Unlike the planetary targets, the Moon is a satellite of a planet, so the Earth, Moon, and Sun must be resolved together at a much smaller physical target radius and integration time step. Lunar impact statistics therefore converge more slowly than planetary Hill-sphere crossing statistics. They may require a separate integration strategy.
\end{itemize}

\section{Conclusion}
\label{sec:conclusion}

We developed a dynamical framework that estimates the intrinsic $\gtrsim 10$~m impact rate on the Solar System planets, tracing from source populations through Hill-sphere close encounters to physical impacts. We applied this framework to size-calibrated populations of NEOs, MBAs, Centaurs, JFCs, and scattering-TNOs over a 300 yr $N$-body integration, and estimated impacts with both a direct periapsis-count method and a population-level B-plane expectation method.

In terms of periapsis-confirmed impactors, Venus records $6^{+3.8}_{-2.7}$ impacts, all from NEOs; Earth has $10^{+11.3}_{-4.9}$, split evenly between NEOs and JFCs; Jupiter receives $2981^{+481}_{-469}$, dominated by JFCs and MBAs; Saturn receives $1228^{+2759}_{-1017}$ from Centaurs; and Mercury, Mars, Uranus, and Neptune have zero predicted direct impacts over the finite integration. The scattering-TNO source population likewise produces no periapsis-confirmed impacts. The $1\sigma$ upper limits for the zero-count cases range from $1.84$ for the terrestrial planets to $1.1\times10^{8}$ for scattering-TNOs onto the outer planets. We also find that the most frequent visitors to a planet are not generally its most frequent impactors.

The two estimators agree for predominantly unbound NEO and JFC encounters but diverge for the MBA–Jupiter case, where the periapsis estimate exceeds the B-plane expectation by a factor of $\sim28$. This difference reveals a distinct Jovian regime dominated partly by slow, bound, repeated-passage trajectories that are not represented by an unbound B-plane calculation. For the Uranus, Neptune, and scattering-TNO channels, zero direct impacts reflect under-sampling of the rare collisions; the B-plane expectation provides the appropriate impact rate.

The main limitations in our work are the source-population normalizations, especially the albedo and size-frequency assumptions used to extend MBA, Centaur, JFC, and scattering-TNO populations to $>10$ m scales and the omission of cometary activity, which may be significant for the predicted JFC impacts over a 300-year integration.

The cometary Earth impactor found in our simulation is not explicitly included in the original Earth decameter gap work \citep{Chow2025decameter}, but doubles the modeled Earth impact rate, suggesting a potential reduction of the decameter gap. In Paper II, we combine the intrinsic impactor population with survey and physical-conversion models and compare the resulting predictions with observed impact rates. We then test whether the Earth source--target mismatch is reduced by the updated modeled rate and whether similar mismatches occur elsewhere in the Solar System.

\section*{Acknowledgment}

We thank all people whose discussions have contributed to this work: Jacob A. Kurlander, for guidance on the MBA and scattering-TNO populations; Pedro H. Bernardinelli, for the regrouped scattering population; Ian Chow, for insightful discussion of the Earth decameter mismatch and guidance on NEOMOD3 sampling; Joseph Murtagh, for guidance on Centaur population generation and discussion on cometary effects; and David Nesvorny, for Centaur population data and guidance on the JFC population generation. We also thank several of these colleagues for generously sharing code to generate these simulation populations, which informed our implementation.

Q.C. and D.S. acknowledge support from the Duke University Trinity College of Arts and Sciences Department of Physics and from the Cosmology Group. D.S. acknowledges support from the Duke University Electrical and Computer Engineering Department.

D.S. is supported by the Department of Energy grant DE-SC0010007, the David and Lucile Packard Foundation, the Templeton Foundation, and the Sloan Foundation.

\appendix
\onecolumngrid
\section{Comparison Between Survey-facing population and Calibrated population}
\label{app:population_comparison}

\begin{figure}[h]
    \centering

    \begin{subfigure}{0.45\textwidth}
        \centering
        \includegraphics[width=\linewidth]{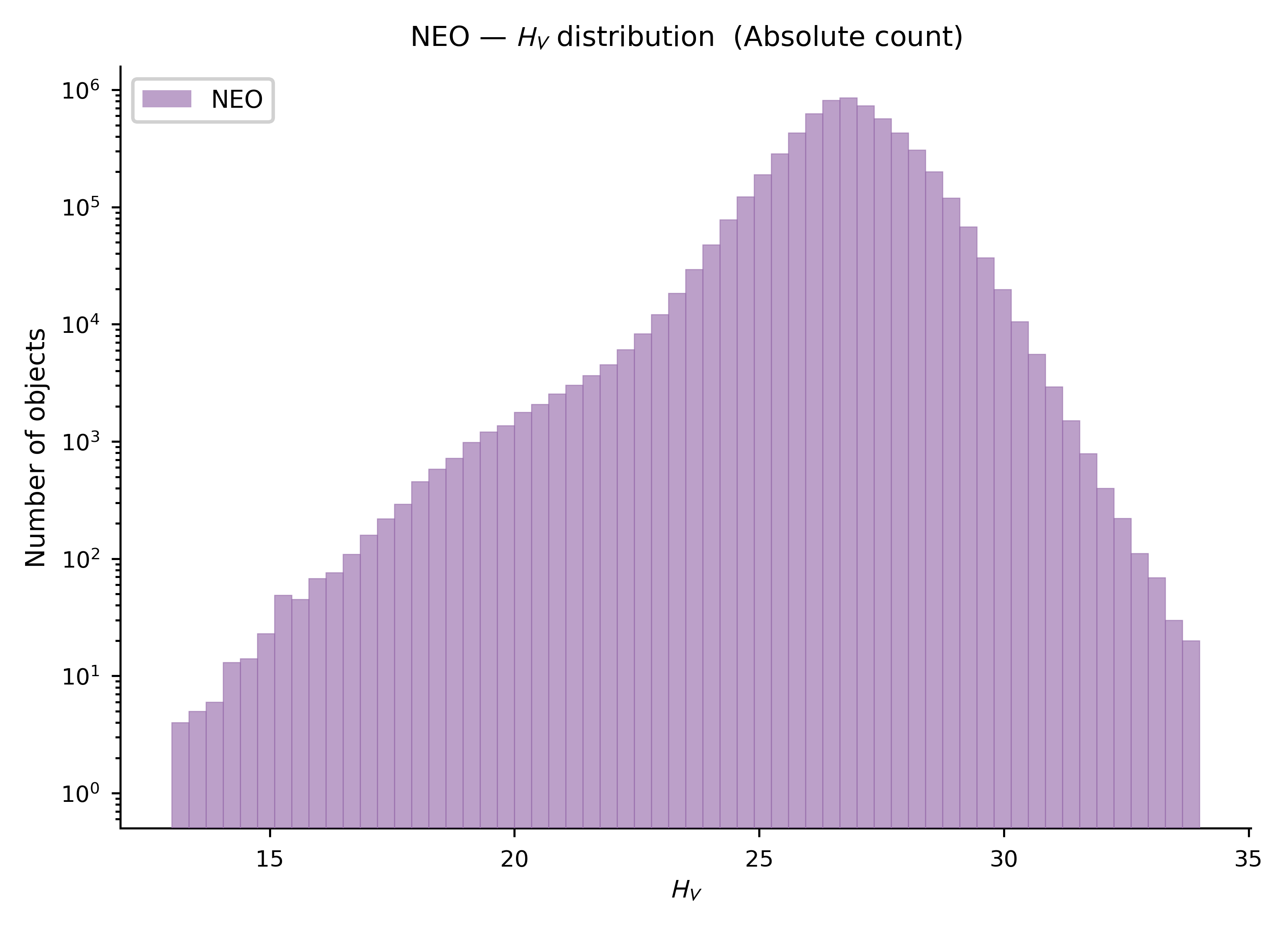}
        \caption{NEO}
        \label{fig:H_theory_NEO}
    \end{subfigure}
    \hfill
    \begin{subfigure}{0.45\textwidth}
        \centering
        \includegraphics[width=\linewidth]{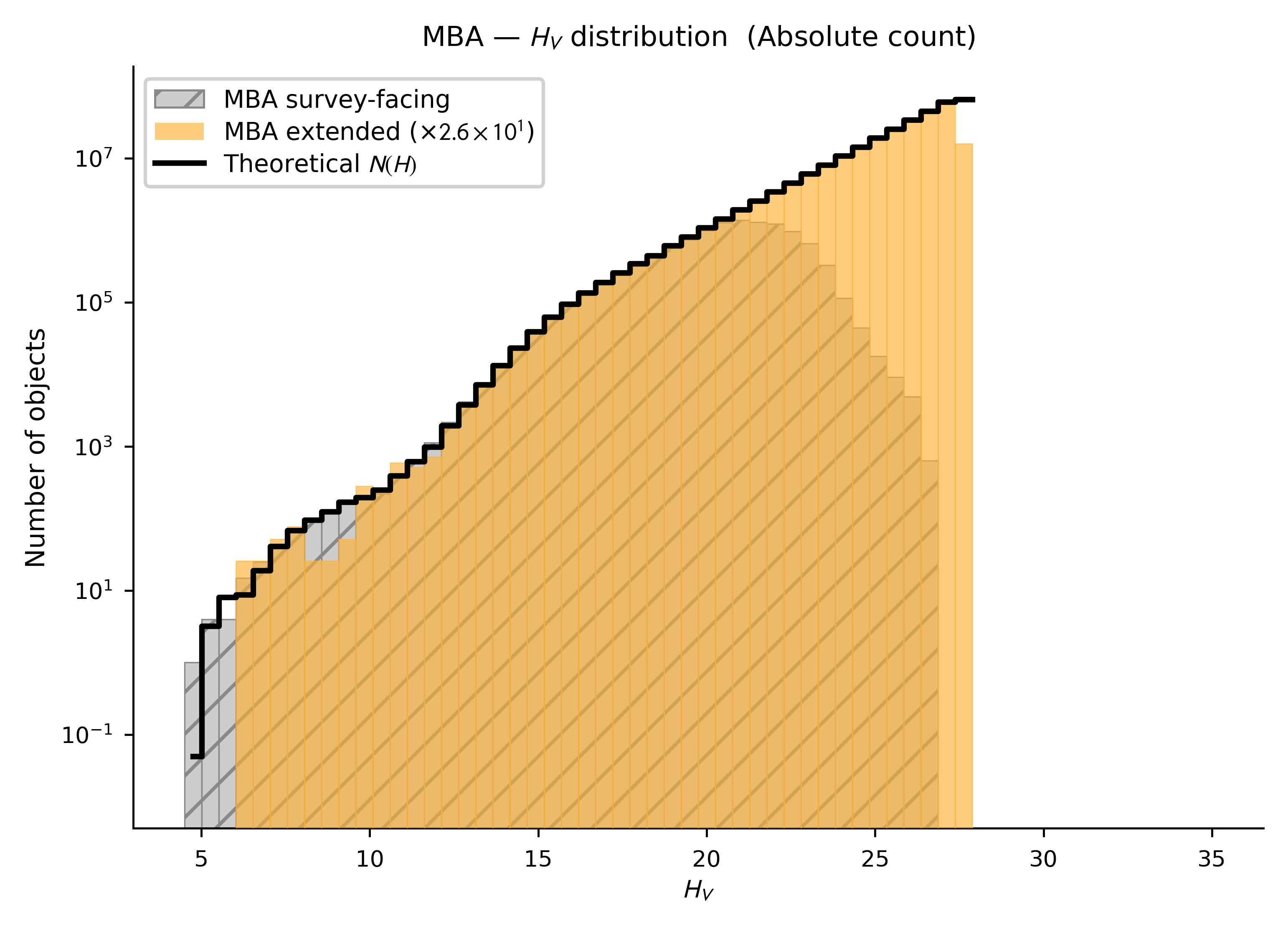}
        \caption{MBA}
        \label{fig:H_theory_MBA}
    \end{subfigure}

    \vspace{1em}

    \begin{subfigure}{0.45\textwidth}
        \centering
        \includegraphics[width=\linewidth]{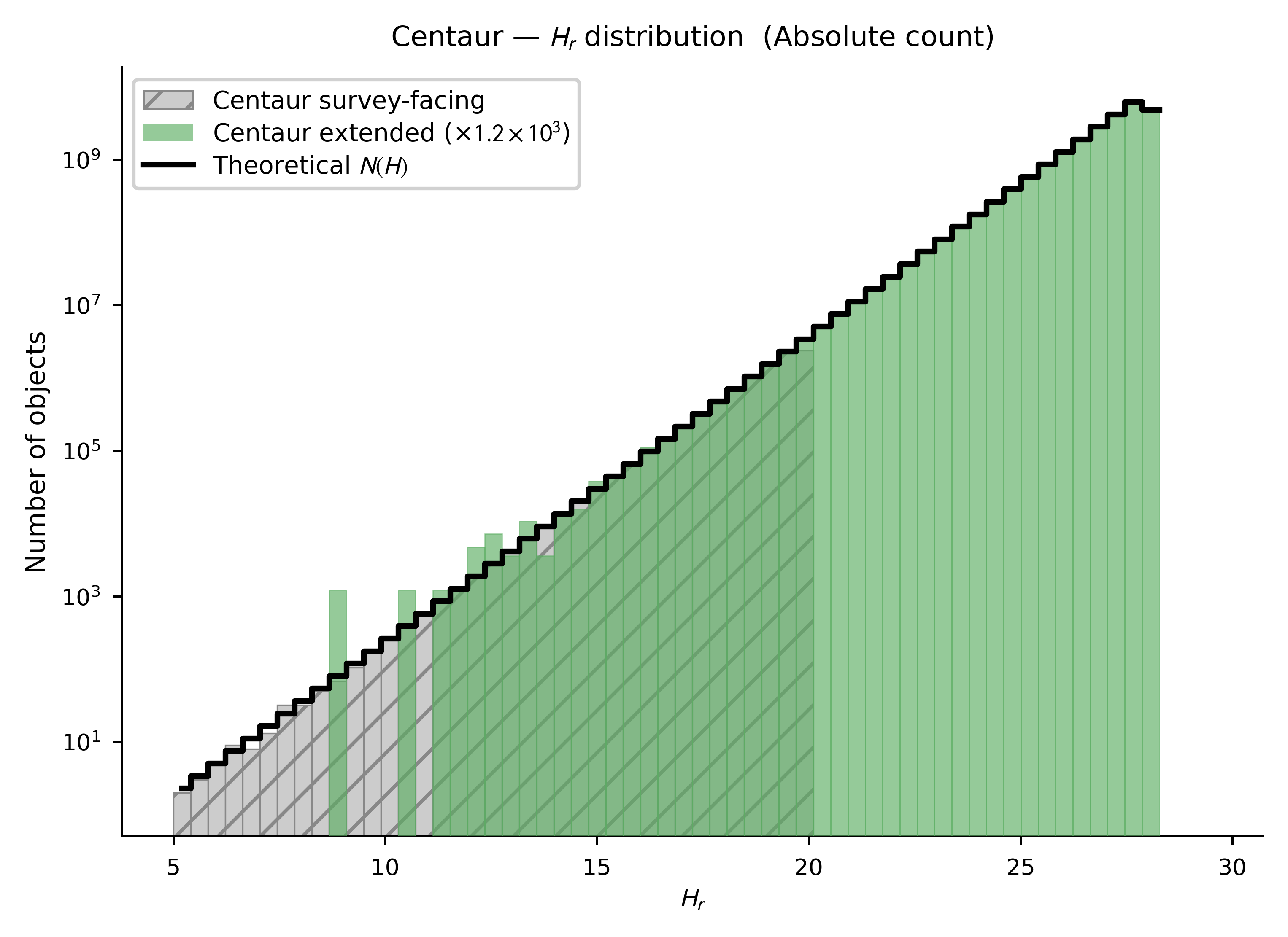}
        \caption{Centaur}
        \label{fig:H_theory_Centaur}
    \end{subfigure}
    \hfill
    \begin{subfigure}{0.45\textwidth}
        \centering
        \includegraphics[width=\linewidth]{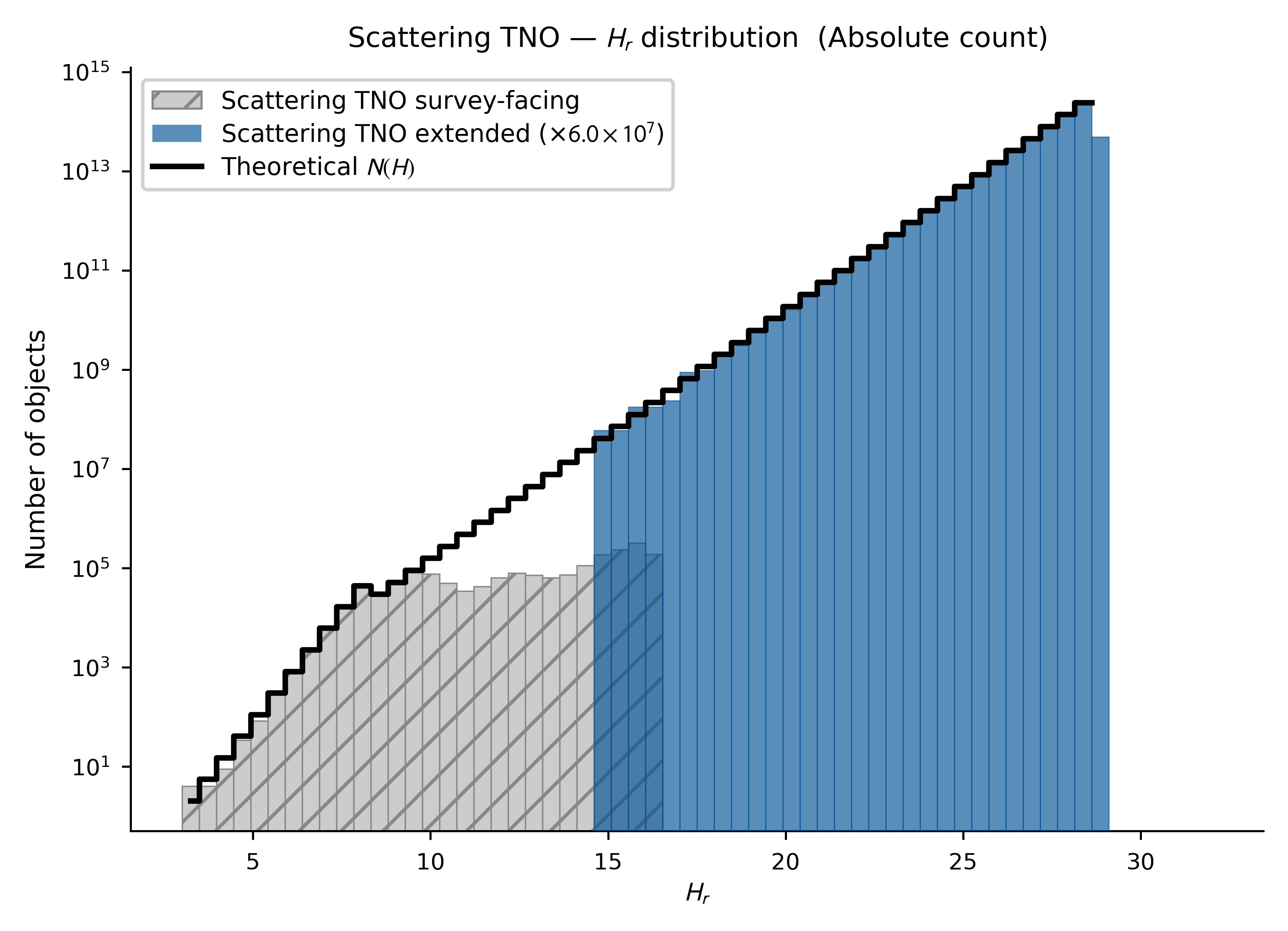}
        \caption{scattering-TNO}
        \label{fig:H_theory_SCO}
    \end{subfigure}
    \vspace{1em}

    \begin{subfigure}{0.45\textwidth}
        \centering
        \includegraphics[width=\linewidth]{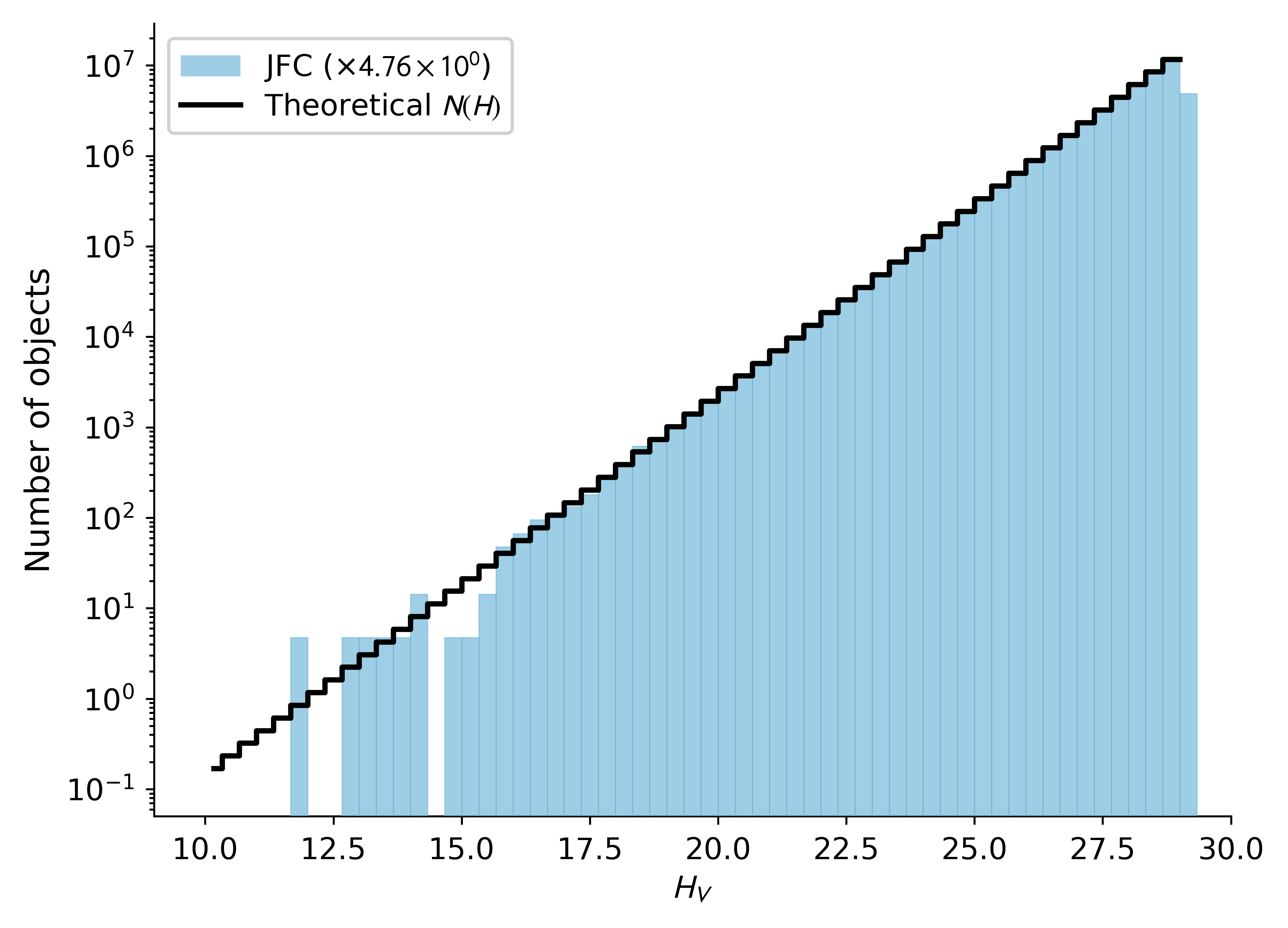}
        \caption{JFC}
        \label{fig:H_theory_JFC}
    \end{subfigure}

    \caption{
    Absolute-magnitude distributions across the five source populations. Each panel compares the survey-facing population (gray hatched histograms), the adopted theoretical cumulative magnitude distribution $N(H)$ (black curves), and the size-extended Monte Carlo sample after corresponding scaling (colored histograms). Panels show (a) NEOs, (b) MBAs, (c) Centaurs, (d) scattering-TNOs, and (e) JFCs. The y-axis gives the absolute number of objects per magnitude bin. For NEOs, the calibrated sample is the same as the survey-facing population, while the MBA, Centaur, and scattering-TNO populations are extended to fainter magnitudes using their respective theoretical magnitude distributions. JFCs do not have a survey-facing population. 
    }
    \label{fig:H_theory_absolute}
\end{figure}

In Figure~\ref{fig:H_theory_absolute}, we compare the theoretical size-extension model with the $H$ distributions of the calibrated samples and the survey-facing populations for all five source groups. They show overall agreement once the calibrated samples are scaled accordingly. The slight misalignment between the calibrated samples and the theoretical model in the bright-end bins, especially for Centaurs and Scattering-TNOs, is due to under-sampling. Both contain relatively few samples (10 million for Centaur, and 20 million for scattering-TNO), in contrast to the rapid rise and large counts at the large $H$ end.

The size-calibrated and survey-facing populations have nearly identical LSST color distributions across the four populations where survey-facing counterparts exist (NEOs, MBAs, Centaurs, and scattering-TNOs). The mean color indices $(u-r)$, $(g-r)$, $(i-r)$, $(z-r)$, and $(y-r)$ agree to within $\lesssim 0.02$ mag in all cases, indicating that the color-assignment procedure samples the same spectral templates independently of H distribution. Thus, the size-calibration step does not introduce a measurable color--observability bias.

The orbital distributions show stronger observational selection effects which are population-dependent. For MBAs, the survey-facing population is shifted toward smaller perihelion distances ($\bar{q}=2.25$ au versus $2.47$ au) and larger eccentricities ($\bar{e}=0.162$ versus $0.134$), consistent with the preferential detection of brighter, more frequently observed inner-belt objects. For scattering-TNOs, the largest offset is in semi-major axis: the extended population has $\bar{a}\approx210$ au ($\sigma\approx124$ au), whereas the survey-facing sample is concentrated near $\bar{a}\approx80$ au ($\sigma\approx85$ au). This reflects the fact that most distant, highly eccentric objects remain below the LSST detection threshold during the survey window. Centaurs show no measurable orbital offset between the two samples ($\Delta\bar{a}<0.01$ au, $\Delta\bar{e}<0.001$), suggesting that the observable Centaur population is a nearly unbiased orbital tracer of the underlying distribution at the simulated brightness limit.

\section{Two-stage Hill-sphere Close Encounters Screening}
\label{app:two stage}
We identify close encounters with a two-stage screening to reduce computation cost, with parameters given in Table~\ref{tab:screening_parameters}. A coarse stage propagates each object at a population-dependent timestep and monitors its distance to each planet against a threshold set to a multiple of the Hill radius ($\alpha_p$). Objects passing the coarse stage are refined in a fine stage that interpolates the stored adaptive output at a smaller timestep, resolving the encounter geometry and recording the planet-relative state vectors needed for the subsequent impact test. Timesteps differ by population to avoid missing brief inner planet encounters while reducing computational cost for less interactive outer planets. 

\begin{deluxetable}{lccccccccccccc}
\tabletypesize{\scriptsize}
\tablewidth{0pt}
\tablecaption{
Two-stage planet close encounters screening settings.
\label{tab:screening_parameters}
}
\tablehead{
\colhead{Pop.} &
\colhead{Years} &
\colhead{$T$} &
\colhead{$\Delta t_{\rm coarse}$} &
\colhead{$\Delta t_{\rm fine}$} &
\colhead{$\Delta t_{\rm half}$} &
\multicolumn{8}{c}{$\alpha_p$} \\
\cline{7-14}
\colhead{} &
\colhead{} &
\colhead{yr} &
\colhead{d} &
\colhead{d} &
\colhead{d} &
\colhead{Merc.} &
\colhead{Ven.} &
\colhead{Earth} &
\colhead{Mars} &
\colhead{Jup.} &
\colhead{Sat.} &
\colhead{Uran.} &
\colhead{Nep.}
}
\startdata
\hline
NEO
& 2025--2325
& 300
& 0.25
& $1/24$
& 1.5
& 4.0
& 1.5
& 1.0
& 1.5
& 1.0
& 1.0
& 1.0
& 1.0 \\
MBA
& 2025--2325
& 300
& 0.25
& $1/24$
& 1.5
& 4.0
& 1.5
& 1.0
& 1.5
& 1.0
& 1.0
& 1.0
& 1.0 \\
JFC
& 2025--2325
& 300
& 0.25
& $1/24$
& 1.5
& 4.0
& 1.5
& 1.0
& 1.5
& 1.0
& 1.0
& 1.0
& 1.0 \\
Centaur
& 2025--2325
& 300
& 5.0
& 1.0
& 90
& 1.0
& 1.0
& 1.0
& 1.0
& 3.0
& 5.0
& 2.0
& 4.0 \\
scattering-TNO
& 2025--2325
& 300
& 5.0
& 1.0
& 90
& 1.0
& 1.0
& 1.0
& 1.0
& 3.0
& 5.0
& 2.0
& 4.0 \\
\enddata
\end{deluxetable}
\onecolumngrid 

\section{Moon-crossing target}
\label{app:moon}

The Moon requires a different integration setup from the planets because it is embedded in the Sun--Earth--Moon system. For planetary targets, the Hill radius is computed with respect to the Sun; for the Moon, the relevant local gravitational sphere is its Hill radius with respect to the Earth, given by
\begin{equation}
  R_{\mathrm{H},\mathrm{Moon}}
  = a_{\mathrm{Moon}\oplus}
    \left(\frac{m_{\mathrm{Moon}}}{3M_\oplus}\right)^{1/3},
  \label{eq:moon_hill_radius_candidate}
\end{equation}
where $a_{\mathrm{Moon}\oplus}$ is the mean Earth--Moon distance, $m_{\mathrm{Moon}}$ is the lunar mass, and $M_\oplus$ is the mass of the Earth. We adopt
\begin{equation}
  R_{\mathrm{H},\mathrm{Moon}}
  \simeq 6.15\times10^{4}~\mathrm{km}
  \simeq 4.11\times10^{-4}~\mathrm{au},
\end{equation}
which corresponds to approximately $35$ lunar radii from the center of the Moon.

The lunar Hill sphere is within Earth's Hill sphere, so we apply the lunar screening only to Earth close encounters. The screening includes a 1-hour coarse pass, which selects candidates approaching within five lunar Hill radii, followed by a 10-minute fine pass that isolates encounters within one Hill radius. We then apply the same periapsis and B-plane impactor estimates used for planets. The resulting lunar encounter and impact statistics are summarized in Table~\ref{tab:moon_master_v2}. Among the 136,753 Earth close encounters (from NEOs and JFCs), 377 ($\sim0.28\%$) passed within the adopted Moon Hill radius. The closest lunar encounter reaches 5,099 km (2.94 lunar radii) from the Moon's center by periapsis distance estimation and an expected number of 0.31 lunar impacts over 300 years. No lunar impacts are identified. The sampling step can be one major limitation. Assuming a mean impact speed of $20\,\mathrm{km\,s^{-1}}$ for Earth impact from our simulation, the 10 min time step implies a travel distance of 6.91 lunar radii. The limitations of the Earth impactor modeling, such as the underestimates of impacts and potential missing cometary-like asteroids, also apply here, since both are drawn from the same underlying populations. 

\begin{table*}[t]
\centering
\caption{Summary of Moon periapsis-method and B-plane results for the 2025--2325 integrations.}
\label{tab:moon_master_v2}
\scriptsize

\begin{tabular}{llrrrrrrrrrrrrr}
\toprule
 & &
\multicolumn{2}{c}{Encounter sample} &
\multicolumn{2}{c}{Periapsis method} &
\multicolumn{2}{c}{B-plane method} &
\multicolumn{4}{c}{Periapsis distance} &
\multicolumn{3}{c}{Encounter speed} \\
\cmidrule(lr){3-4}
\cmidrule(lr){5-6}
\cmidrule(lr){7-8}
\cmidrule(lr){9-12}
\cmidrule(lr){13-15}
Group &
Planet &
$N_\mathrm{cross}$ &
$N_\mathrm{unb}$ &
$N_\mathrm{imp}$ &
$f_\mathrm{imp}$ &
$\bar{f}_{\rm B}$ &
$\langle N_{\rm imp}\rangle_{\rm B}$ &
$r_{peri,\min}$ &
$r_{peri,10}$ &
$\bar{r}_{peri}$ &
$r_{peri,90}$ &
$v_{\min}$ &
$\bar{v}$ &
$v_{\max}$ \\
 & & & & & $(10^{-3})$ & $(10^{-3})$ & &
\multicolumn{4}{c}{$(R_{\rm Moon})$} &
\multicolumn{3}{c}{(km\,s$^{-1}$)} \\
\midrule
NEOs\tablenotemark{a} &
Moon &
\num{367} & \num{367} & 0 & 0.0000 & 0.83329 & 0.3058 &
2.94 & 11.3 & 24.1 & 33.5 &
2.48 & 14.15 & 46.35 \\
JFCs\tablenotemark{b} &
Moon &
\num{10} & \num{10} & 0 & 0.0000 & 0.80724 & 0.0077 &
10.62 & 12.4 & 19.3 & 26.2 &
15.24 & 20.61 & 25.99 \\
\bottomrule
\end{tabular}%

\raggedright
\tablecomments{
$N_\mathrm{cross}$ is the number of objects entering Moon Hill radius,
$N_\mathrm{unb}$ is the number of unbound encounters with $\varepsilon>0$,
$N_\mathrm{imp}$ is the number of periapsis impactors with $r_{peri} \le R_{\rm Moon}$,
and $f_\mathrm{imp}=N_\mathrm{imp}/N_\mathrm{cross}$.
$\bar{f}_{\rm B}$ is the mean B-plane impact probability over unbound crossers, and
$\langle N_{\rm imp}\rangle_{\rm B}= \sum f_{\rm B}$ is the expected number of impacts.
Periapsis distances are in units of the Moon's equatorial radius,
$R_{\rm Moon}=1737.4\,\mathrm{km}$.}
\tablenotetext{a}{Raw simulation counts presented; scaling factor $f=1$.}
\tablenotetext{b}{Counts scaled by $f=4.756$; the factor multiplies $N_\mathrm{cross}$, $N_\mathrm{unb}$, $N_\mathrm{imp}$, and $\langle N_{\rm imp}\rangle_{\rm B}$; ratios and means are unaffected.}
\end{table*}

\section{Validation}
\label{app:validation}
\paragraph{Validation of closest-approach distance with JPL Horizons}
We compare a subset of our closest-approach distance results against independent ephemerides from JPL Horizons \citep{Giorgini1996}, with close agreement. We choose the survey-facing MBA population as the validation sample, because it produces a large number of Mars and Jupiter Hill-sphere encounters, representative of both strong and weak gravitational trapping cases. For each object, we submit the same initial orbital elements over the same time period to JPL Horizons and calculate the periapsis distance by the same position-vector difference. The two estimates agree closely. The median difference is $1.43\times10^{-6}$ au for Jupiter and $3.19\times10^{-9}$ au for Mars, while 98.5\% of Jupiter pairs and 100\% of Mars pairs agree to within $10^{-3}$ au. 



The Hill-sphere classifications are also nearly identical. All 8,917 Jupiter Hill-sphere close encounters flagged by our pipeline are confirmed by Horizons. For Mars, 596 of 597 close encounters agree. The single discrepant case is a boundary event, where the periapsis estimate from our pipeline lies just inside $R_{\rm H,Mars}=0.007$ au, while the Horizons sampled minimum distance is $0.007014$ au, falling just outside the threshold. 

\paragraph{Validation of impactor identification with Adaptive-IAS15 integrations}
\label{sec:v5_adaptive_validation}

We compare our pipeline with a pure adaptive IAS15 run over the same 300 yr NEO window to validate the impactor-classification results from our pipeline. The adaptive run launches the IAS15 integration (minimum timestep $10^{-6}$ d $\approx 86$ ms) and records a direct impact when $d_{peri}(t) \le R_{\mathrm{th},p}$. We use the NEO population for this validation because it produces close approaches at the largest number of planets, while the number of objects remains computationally manageable. Our pipeline recovers all 13 adaptive-IAS15 impacts (recall $1.00$) out of a total 16 predicted impacts (precision $\simeq0.81$). The three false positives are Venus impactors, while Earth and Jupiter give 100\% precision. The $r_{peri}$ and $d_{\mathrm{IAS15}}$ minimum distances agree within 7\% ($<0.1\%$ for Jupiter, $6.9\%$ for Earth).


\section{Simulation Run Time}
\label{app:simulation run time}

All $N$-body integrations were executed as \texttt{SLURM} array jobs on the Duke Computing Cluster, with single-core array tasks each integrated 500--1000 objects (up to 400 concurrent tasks) for a total of $9\times10^{7}$ objects simulated. Typical task-level wall times were 18--20 min for Centaur and scattering-object chunks and 1.5--4.8 hr for NEO, MBA and JFC chunks. The subsequent periapsis and B-plane post-processing is computationally negligible by comparison, completing in 3--4 s for typical NEO files of size $\sim 0.3$ GB and in less than 5 min for the largest Centaur files of size $\sim20$ GB. 



\section{Survey Facing Population Results}
\label{app:original_population} 

This appendix shows the same diagnostics as the main Results section, but for the survey-facing population with the detection-limit cut. Major findings are in Table~\ref{tab:s3m_impactors}. 

The survey-facing and calibrated (intrinsic) populations agree at the $\lesssim 30\%$ level in per-close-encounter impact rate for Centaurs and scattering-TNOs at Saturn and Uranus, where the survey-selection function does not strongly correlate the size distribution with the orbital parameter distribution. The principal exception is scattering-TNO encounters at Neptune, where the survey-facing catalog underrepresents slow long-period scatterers that carry large B-plane weight. For MBAs the discrepancy is larger and qualitative. The survey-facing catalog is depleted in the slow outer-belt close encounters that dominate the calibrated Monte Carlo, leading to a factor of $\sim2.75$ difference in raw periapsis impact fraction and a factor of $\sim8$ difference in the scaled Centaur impactor count once the bound close encounter caveat is applied. In both cases the B-plane method over unbound close encounters is the more reliable estimator; the quoted results in the body of the paper use the calibrated extended population throughout, with the survey-facing rates serving as a consistency check and an upper limit on the observationally accessible population.

\begin{table}[htbp]
  \centering
  \caption{%
    Survey-facing populations 300-year periapsis and B-plane results.
  }
  \label{tab:s3m_impactors}
  \setlength{\tabcolsep}{5pt}
  \begin{tabular}{llrrrrr}
  \toprule
  Population & Planet
    & $N_\mathrm{cross}$ & $N_\mathrm{unb}$
    & $N_\mathrm{imp}$ (bd/unb)
    & $\langle N_{\rm imp}\rangle_{\rm B}$
    & $\bar{f}_\mathrm{B}\ (\times10^{-3})$ \\
  \midrule
  MBAs   & Jupiter & 8\,917 & 6\,450  & \textbf{19} (16/3) & 4.34  & 0.529 \\
            & Mars    &    597  &   597   & 0 (0/0)           & 0.009 & 0.014 \\
            & Saturn  &    239  &   238   & 0 (0/0)           & 0.061 & 0.243 \\
  \midrule
  Centaurs & Saturn  &  5\,778 &  5\,754 & \textbf{1} (0/1) & 0.476 & 0.083 \\
              & Neptune & 22\,335 & 22\,304 & 0 (0/0)          & 0.180 & 0.008 \\
              & Uranus  & 23\,287 & 23\,272 & 0 (0/0)          & 0.255 & 0.011 \\
              & Jupiter &       5 &       5 & 0 (0/0)          & 0.002 & 0.422 \\
  \midrule
  scattering-TNOs & Saturn  & 4\,275 & 4\,275 & 0 (0/0) & 0.242 & 0.057 \\
          & Neptune & 1\,476 & 1\,476 & 0 (0/0) & 0.004 & 0.003 \\
          & Uranus  & 1\,304 & 1\,304 & 0 (0/0) & 0.004 & 0.003 \\
  \bottomrule
  \end{tabular}
  \raggedright
  \footnotesize
  \tablecomments{$N_\mathrm{cross}$: close encounters within one Hill radius; $N_\mathrm{unb}$: unbound objects ($\varepsilon > 0$); $N_\mathrm{imp}$: periapsis impactors ($r_p \le R_\mathrm{p}$), with bound/unbound breakdown in parentheses; $\langle N_{\rm imp}\rangle_{\rm B}$: expected impact count from B-plane method (unbound close encounters only); $\bar{f}_\mathrm{B}$: mean B-plane probability per unbound close encounter.}
\end{table}

\begin{table}
\centering
\caption{Orbital elements and Jupiter Tisserand parameters of the integration-limited objects. \label{tab:s100brmpa}}
\begin{tabular}{lcccccc}
\hline
ObjID & $a$ (au) & $e$ & $i$ (deg) & $q$ (au) & $Q$ (au) & $T_{\rm J}$ \\
\hline
j590504   & 3.978 & 0.396 & 0.76 & 2.40  & 5.55 & 2.91 \\
j4738535  & 2.906 & 0.799 & 3.35 & 0.58  & 5.23 & 2.69 \\
j7293058  & 3.804 & 0.352 & 3.16 & 2.46  & 5.14 & 2.97 \\
j7421066  & 3.593 & 0.430 & 1.60 & 2.05  & 5.14 & 2.95 \\
j8768360  & 3.462 & 0.464 & 5.02 & 1.86  & 5.07 & 2.94 \\
j8855136  & 3.593 & 0.430 & 1.60 & 2.05  & 5.14 & 2.95 \\
j9417471  & 3.658 & 0.401 & 4.20 & 2.19  & 5.12 & 2.95 \\
\hline
S100bRmPa & 3.980 & 0.285 & 0.69 & 2.843 & 5.11 & 2.98 \\
\hline
\end{tabular}
\end{table}

\section{Note on the excluded co-planar objects}
\label{app:s100brmpa}
The high-accuracy adaptive integration of eight co-planar objects with Jupiter (one from MBA and 7 from JFC) are computationally expensive and is excluded in our work. Their orbital parameters are summarized in Table~\ref{tab:s100brmpa}. One example is \texttt{S100bRmPa} from MBA, whose orbit is $q=2.843$ au, $e=0.285$, $i=0.69^\circ$, $a=3.98$ au, and $Q=5.11$ au. Its Tisserand parameter with respect to Jupiter, $T_{\rm J}\simeq2.98$, places it near the conventional boundary between asteroidal and JFC-like dynamics. Because its aphelion approaches Jupiter's orbital region and its inclination is nearly coplanar with Jupiter, the object encounters Jupiter deeply and repeatedly. During such encounters, an adaptive integrator such as IAS15 reduces its internal time step to maintain the specified error tolerance \citep{rein2015ias15}, and the cumulative step count becomes impractical over the full 300-year integration window (exceeding 48 hr per object; the 48 hr limit is simply the longest wall-clock time we tested, not a physical or algorithmic ceiling). We therefore document the exclusion explicitly and retain the object's orbital elements and computational history for future targeted integrations.




\bibliography{ref}{}
\bibliographystyle{aasjournalv7}

\end{document}

%% file: input_population_sum.tex
\begin{table*}[ht!]
\centering
\caption{Summary of input small-body populations, magnitude ranges, and corresponding literature used in this work.}
\label{tab:population_summary}
\setlength{\tabcolsep}{5pt}
\footnotesize
\begin{tabular}{llccc}
\toprule
Population & Model & $H$ range & $N_{\mathrm{sim}}$ & Reference \\
\midrule
NEO & Survey-facing/extended & $H_V < 34.81$ & $6.06\times10^{6}$ & \cite{Kurlander2025,NesvornyNEOMOD3} \\
\midrule
\multirow{2}{*}{MBA}
 & Survey-facing & $H_V < 26.69$ & $1.11\times10^{7}$ & \multirow{2}{*}{\cite{Kurlander2025,Grav2011}} \\
 & Extended      & $H_V < 27.80$ & $2.57\times10^{8}$ & \\
\midrule
\multirow{2}{*}{Centaur}
 & Survey-facing & $H_r < 24.5$  & $9.47\times10^{6}$  & \multirow{2}{*}{\cite{Murtagh2025,Nesvorny2019}} \\
 & Extended      & $H_r < 28.10$ & $2.39\times10^{10}$ & \\
\midrule
JFC & Extended & $H_V < 29.11$ & $4.76\times10^{7}$ & \cite{Nesvorny2017,Murtagh2025} \\
\midrule
\multirow{2}{*}{Scattering TNO}
 & Survey-facing & $H_r < 16.38$ & $1.82\times10^{6}$  & \multirow{2}{*}{\cite{Kurlander2025,lawler2018ossos}} \\
 & Extended      & $H_r < 28.66$ & $5.95\times10^{14}$ & \\
\bottomrule
\end{tabular}
\vspace{2mm}
\tablecomments{Extended populations represent the intrinsic $D>10$~m population, whereas survey-facing populations include the LSST detectability pre-cut. $N_{\mathrm{sim}}$ denotes the number of objects corresponding to each population. Note that we did not simulate the full extended populations, as detailed in Section~\ref{sec:method_calibrated_population}.}
\end{table*}

%% file: master_v2_longrotatetable-paper.tex


\begingroup
\sisetup{group-separator={,}, group-minimum-digits=4}
\begin{deluxetable*}{llrrrrrrrrrrrrr}
\tabletypesize{\scriptsize}
\tablewidth{0pt}
\setlength{\tabcolsep}{2pt}
\renewcommand{\arraystretch}{1.4}

\tablecaption{Summary of periapsis-method and B-plane results for the 2025--2325 integrations.\label{tab:master_v2}}

\tablehead{
\colhead{} &
\colhead{} &
\multicolumn{2}{c}{Encounter sample} &
\multicolumn{2}{c}{Periapsis method} &
\multicolumn{2}{c}{B-plane method} &
\multicolumn{4}{c}{Periapsis distance} &
\multicolumn{3}{c}{Encounter speed} \\
\colhead{Group} &
\colhead{Planet} &
\colhead{$N_\mathrm{cross}$} &
\colhead{$N_\mathrm{unb}$} &
\colhead{$N_\mathrm{imp}$} &
\colhead{$f_\mathrm{imp}$ ($10^{-3}$)} &
\colhead{$\bar{f}_{\mathrm{B\mbox{-}plane}}$ ($10^{-3}$)} &
\colhead{$\langle N_{\rm imp}\rangle_{\rm B}$} &
\colhead{$r_{p,\min}$} &
\colhead{$r_{p,10}$} &
\colhead{$\bar{r}_p$} &
\colhead{$r_{p,90}$} &
\colhead{$v_{\min}$} &
\colhead{$\bar{v}$} &
\colhead{$v_{\max}$}
}
\startdata
 & & & & & & & & \multicolumn{4}{c}{($R_p$)} & \multicolumn{3}{c}{(km\,s$^{-1}$)} \\
\tableline
\shortstack[l]{NEOs\tablenotemark{a}}
  & Earth
        & \num{120093} & \num{120091} & \textbf{5}  & 0.0416 & 0.04578 & 5.498
        &  0.569 &  61.7 & 146.5 & 219.5 &  0.99 & 14.74 &  49.03 \\
 & Jupiter
        & \num{30062}  & \num{30062}  & \textbf{11} & 0.3659 & 0.2375 & 7.140
        &  0.167 & 140.8 & 438.7 & 693.9 &  3.90 &  8.64 &  78.13 \\
 & Venus
        & \num{42485}  & \num{42484}  & \textbf{6}  & 0.1412 &  0.0726 & 3.084
        &  0.040 &  45.8 & 104.9 & 155.0 &  1.29 & 17.59 &  57.31 \\
 & Mars
        & \num{34006}  & \num{34006}  & 0           & 0.000  & 0.0133 & 0.453
        &  1.09  &  94.8 & 203.9 & 292.1 &  1.39 & 13.14 &  39.88 \\
 & Mercury
        &     387 &     387 & 0           & 0.000  & 0.2048 & 0.079
        &  2.69  &  21.5 &  47.4 &  68.7 &  7.38 & 23.49 &  52.72 \\
 &Saturn
         &       8 &       8 & 0           & 0.000  & 0.04293 & 0.0003
        & 221.1  & 278.1 & 687.7 & \num{1015} &  3.51 &  5.97 &   7.24 \\
\tableline
\shortstack[l]{MBAs\tablenotemark{b}}
  & Jupiter
    & \num{273121} & \num{57892} & \textbf{\num{1334}} & 4.884 & 0.8280 & 47.94
    & 0.001  &  19.8 & 262.3 & 641.8 & 2.20 &  4.03 &  31.86 \\
 & Mars
  & \num{2180}   & \num{2180}  & 0 & 0.000 & 0.01311 & 0.0286
& 34.7 & 99.4 & 204.1 & 286.4 & 6.77 & 11.62 &  18.91 \\
 & Saturn
  & \num{8362}   & \num{8311}  & 0 & 0.000 & 0.2574 & 2.140
  & 10.5 & 237.8 & 633.0 & 958.7 & 1.22 & 2.86 & 11.01 \\
\tableline
\shortstack[l]{JFCs\tablenotemark{c}}
    & Jupiter
        & \num{6644622} & \num{6638805} & \textbf{\num{1636}} & 0.2462 & 0.26991 & 1791.86
        &  0.001 & 150.3 & 432.8 & 686.8 &  2.23 &  8.80 & 92.24 \\
    & Saturn
        & \num{374464}  & \num{374464}  & \textbf{33}          & 0.0889 & 0.04344 &  16.27
        &  0.032 & 290.8 & 658.4 & 963.8 &  1.53 &  6.51 & 59.64 \\
    & Earth
        & \num{16660}   & \num{16660}   & \textbf{5}           & 0.2855 & 0.02803 &   0.46
        &  0.426 &  61.9 & 138.5 & 209.1 &  8.37 & 18.59 & 47.24 \\
    & Mars
        & \num{6530}    & \num{6530}    & 0                    & 0.000  & 0.01223 &   0.08
        & 12.30  &  80.8 & 181.8 & 272.7 &  5.11 & 15.07 & 34.70 \\
    & Venus
        & \num{3833}    & \num{3833}    & 0                    & 0.000  & 0.04804 &   0.18
        &  6.891 &  43.8 &  97.5 & 148.1 & 11.13 & 22.14 & 53.41 \\
\tableline
\shortstack[l]{Centaurs\tablenotemark{d}}
  & Saturn
    & \num{11784289} & \num{11784289} & \textbf{\num{1195}} & 0.1014 & 0.06206 & 731.31
    &  0.25 & 294.4 & 669.4 & 967.6 & 1.44 &  7.64 & 23.68 \\
 & Neptune
    & \num{56251728} & \num{56183611} & 0 & 0.000 & 0.00780 & 438.50
    &  1.38 & \num{1415} & \num{3069} & \num{4405} & 0.35 &  2.65 & 10.01 \\
 & Uranus
    & \num{57471864} & \num{57446768} & 0 & 0.000 & 0.01082 & 621.36
    &  1.90 & 839.1 & \num{1816} & \num{2607} & 0.44 &  3.58 & 14.29 \\
 & Jupiter
    & \num{1195} & \num{1195} & 0 & 0.000 & 0.61086  & 0.73
    & 139.8 & 139.8 & 139.8 & 139.8 & 7.43 &  7.43 &  7.43 \\
\tableline
\shortstack[l]{Scattering TNOs\tablenotemark{e}}
  & Neptune
  & $1.07{\times}10^{11}$ & $1.08{\times}10^{11}$ & 0 & 0.000 & 0.01341 & \num{1429383}
  & 13.98   & \num{1344} & \num{3056} & \num{4412} & 0.48 &  2.17 &  6.13 \\
 & Saturn
  & $9.25{\times}10^{9}$ & $8.87{\times}10^{9}$ & 0 & 0.000 &  0.06290 & \num{581981}
  & 66.8   & 320.3 & 673.0 & 959.5 & 3.00 &  5.02 &  7.64 \\
 & Uranus
  & $2.17{\times}10^{10}$ & $2.20{\times}10^{10}$ & 0 & 0.000 & 0.00702 & \num{152153}
  & 57.1 & 858.8 & \num{1767} & \num{2552} & 1.33 &  3.51 &  7.03 \\
\enddata

\tablecomments{$N_\mathrm{cross}$ is the number of objects entering one Hill radius, $N_\mathrm{unb}$ is the number of unbound encounters with $\varepsilon>0$, $N_\mathrm{imp}$ is the number of periapsis impactors with $r_p \le R_\mathrm{planet}$, and $f_\mathrm{imp}=N_\mathrm{imp}/N_\mathrm{cross}$. $\bar{f}_{\mathrm{B\mbox{-}plane}}$ is the mean B-plane impact probability over unbound crossers, and $\langle N_{\rm imp}\rangle_{\rm B}= \sum f_{\mathrm{B\mbox{-}plane}}$ is the expected number of impacts from the unbound crosser ensemble.}
\tablenotetext{a}{Raw simulation counts presented; scaling factor $f=1$, as the NEO sample already matches the true population size.}
\tablenotetext{b}{Counts scaled by $f=25.65$, the ratio of the true population to the simulated sample. The factor multiplies $N_\mathrm{cross}$, $N_\mathrm{unb}$, $N_\mathrm{imp}$, and $\langle N_{\rm imp}\rangle_{\rm B}$; ratios and means are unaffected.}
\tablenotetext{c}{Counts scaled by $f=4.756$, as in note~(b).}
\tablenotetext{d}{Counts scaled by $f=1195.04$, as in note~(b).}
\tablenotetext{e}{Counts scaled by $f=2.975\times10^{7}$, as in note~(b).}

\end{deluxetable*}
\endgroup

%% file: uncertainty.tex
\begingroup
\sisetup{group-separator={,}, group-minimum-digits=4}
\begin{deluxetable*}{ll rr rc rc | rr rc rc}
\tabletypesize{\scriptsize}
\tablewidth{0pt}
\setlength{\tabcolsep}{1pt}
\renewcommand{\arraystretch}{1.4}

\tablecaption{Counting-only and combined counting-plus-scaling ($\sigma_f=0.2$) uncertainties for the periapsis and B-plane impact estimates.\label{tab:unc_combined_full}}

\tablehead{
\colhead{} & \colhead{} &
\multicolumn{6}{c|}{Periapsis method} &
\multicolumn{6}{c}{B-plane method} \\
\cline{3-8}\cline{9-14}
\colhead{} & \colhead{} &
\colhead{} & \colhead{} &
\multicolumn{2}{c}{raw count} &
\multicolumn{2}{c|}{scaled, incl.\ $20\%\,\sigma_f$} &
\colhead{} & \colhead{} &
\multicolumn{2}{c}{raw count} &
\multicolumn{2}{c}{scaled, incl.\ $20\%\,\sigma_f$} \\
\colhead{Group} & \colhead{Planet} &
\colhead{$N_{\rm raw}$} & \colhead{$N_{\rm imp}$} &
\colhead{$\epsilon$} & \colhead{$1\sigma$ range} &
\colhead{$\epsilon$} & \multicolumn{1}{c|}{$1\sigma$ range}  &
\colhead{$N_{B,{\rm raw}}$} & \colhead{$\langle N\rangle_{\rm B}$} &
\colhead{$\epsilon$} & \colhead{$1\sigma$ range} &
\colhead{$\epsilon$} & \colhead{$1\sigma$ range}
}
\startdata
\shortstack[l]{NEO\tablenotemark{a}}
  & Earth   &  5 &          5 &  45\% & 2.84--8.38 &  49\% & 2.62--8.53            &       5.498 &              5.498 &   43\% & 3.15--7.84 &   47\% & 2.91--8.09 \\
  & Jupiter & 11 &         11 &  30\% & 7.73--15.40 &  36\% & 7.06--15.90            &      7.140 &              7.140 &
  37\% & 4.47--9.81  &   42\% & 4.11--10.2 \\
  & Venus   &  6 &          6 &  41\% & 3.62--9.58 &  45\% & 3.33--9.78            &      3.084 &              3.084 &
  57\% & 1.33--4.84 &  60\% & 1.22--4.94 \\
  & Mars    &  0 &          0 & \nodata & $<$1.84 & \nodata & $<$1.84               &     0.453 &             0.453 &  149\% & 0--1.13    &  150\% & 0--1.13\\
  & Mercury &  0 &          0 & \nodata & $<$1.84 & \nodata & $<$1.84               &    0.079 &            0.079 &  355\% & 0--0.361    &  356\% & 0--0.361\\
& Saturn  &  0 &          0 & \nodata & $<$1.84 & \nodata & $<$1.84               &    0.0003 &            0.0003 &
  \num{5399}\% & 0--0.019    & \num{5399}\% & 0--0.019 \\
\tableline
\shortstack[l]{MBA \tablenotemark{b}}
  & Jupiter & 52 & \num{1334} &  14\% & 44.8--60.3 &  24\% & \num{1010}--\num{1674} &      1.87 &              47.9 &   73\% & 0.502--3.24 &   76\% & 11.6--84.3 \\
  & Mars    &  0 &          0 & \nodata & $<$1.84 & \nodata & $<$47.2               &   0.00111 &            0.0286 & \num{3000}\% & 0--0.0345 & \num{3000}\% & 0--0.885 \\
  & Saturn  &  0 &          0 & \nodata & $<$1.84 & \nodata & $<$47.2               &    0.0834 &              2.14 &  346\% & 0--0.372    &  347\% & 0--9.56 \\
\tableline
\shortstack[l]{JFC\tablenotemark{c}}
    & Jupiter & 344 & \num{1636} &  5\% & 325--364   &  21\% & \num{1297}--\num{1976} &     376.76 &  \num{1791.86} &    5.15\% &
  357.35--396.17    &   20.65\% & \num{1421.79}--\num{2161.93} \\
    & Saturn  &  7 &         33 &  38\% & 4.42--10.8 &  43\% & 19.3--52.4            &      3.4201 &              16.27 &   54.07\% &
  1.571--5.269  &   57.65\% & 6.890--25.650 \\
  & Earth   &  1 &          5 & 100\% & 0.173--3.30 & 102\% & 0.708--15.7          &    0.0967 &              0.46 &  322\% & 0--0.408    &  322\% & 0--1.94 \\
  & Mars    &  0 &          0 & \nodata & $<$1.84 & \nodata & $<$8.76               &    0.0168 &              0.08 &  771\% & 0--0.147    &  771\% & 0--0.697 \\
  & Venus   &  0 &          0 & \nodata & $<$1.84 & \nodata & $<$8.76               &    0.0379 &              0.18 &  514\% & 0--0.232    &  514\% & 0--1.11 \\
\tableline
\shortstack[l]{Centaur \tablenotemark{d}}
    & Saturn  &  1 & \num{1195} & 100\% & 0.173--3.30 & 102\% & 178--\num{3954}      &     0.6120 &               731.31 &  127.83\% &
  0--1.394     &  129.38\% & 0--\num{1677.5} \\
    & Neptune &  0 &          0 & \nodata & $<$1.84 & \nodata & $<$\num{2200}         &     0.3670 &               438.50 &  165.07\% &
  0--0.9728    &  166.28\% & 0--\num{1167.62} \\
    & Uranus  &  0 &          0 & \nodata & $<$1.84 & \nodata & $<$\num{2200}         &     0.5199 &               621.36 &  138.69\% &
  0--1.241     &  140.12\% & 0--\num{1492.03} \\
    & Jupiter &  0 &          0 & \nodata & $<$1.84 & \nodata & $<$\num{2200}         & 0.000611 &               0.73 & \num{4045.57}\%
  & 0--0.02533 & \num{4045.62}\% & 0--30.263 \\
\tableline
\shortstack[l]{Scattering \tablenotemark{e} \\TNOs}
    & Neptune &  0 &          0 & \nodata & $<$1.84 & \nodata & $<1.10\times10^{8}$ &    0.048 & $1.43\times10^{6}$ &
  \num{456}\% & 0--0.267    &   \num{457}\% & 0--$7.96\times10^{6}$ \\
    & Saturn  &  0 &          0 & \nodata & $<$1.84 & \nodata & $<1.10\times10^{8}$ &    0.0196& $5.82\times10^{5}$ &
  \num{715}\% & 0--0.159 & \num{715}\% & 0--$4.74\times10^{6}$ \\
    & Uranus  &  0 &          0 & \nodata & $<$1.84 & \nodata & $<1.10\times10^{8}$ &   0.0051 & $1.52\times10^{5}$ &
  \num{1398}\% & 0--0.077 & \num{1398}\% & 0--$2.28\times10^{6}$ \\
 \enddata

\tablecomments{
$N_{\rm raw}$ and $N_{B,{\rm raw}}=\sum_i p_i$ are the raw (unscaled) periapsis count and B-plane expected count; $N_{\rm imp}=f\,N_{\rm raw}$ and $\langle N\rangle_{\rm B}=f\,N_{B,{\rm raw}}$ are the scaled values in Table~\ref{tab:master_v2}, with normalization factors $f$ listed in notes~a--d. \\
The counting-only columns give the Poisson uncertainty of the raw count alone, fractional uncertainty $\epsilon=1/\sqrt{N_{\rm raw}}$, with the $1\sigma$ range reported in raw, unscaled units \citep{Gehrels1986}. \\
For periapsis method, uncertainty ranges are exact $68.3\%$ Poisson intervals, and null channels give a one-sided raw upper limit of $1.841$ events.\\
For B-plane method, the counting uncertainty is the Poisson-binomial $\sqrt{\sum_i p_i(1-p_i)}\simeq\sqrt{N_{B,{\rm raw}}}$, with ranges symmetric and clipped at zero.\\
The ``incl.\ $20\%\,f$'' columns for both methods add the normalization uncertainty $\epsilon_f=0.2$ in quadrature, $\epsilon=\sqrt{1/N_{\rm raw}+\epsilon_f^2}$, with the $1\sigma$ range on the scaled impact number ($\epsilon_f N$ added in quadrature to each side of the interval). }
\tablenotetext{a}{Unscaled; $f=1$.}
\tablenotetext{b}{$f=25.65$.}
\tablenotetext{c}{$f=4.756$.}
\tablenotetext{d}{$f=1195.04$.}
\tablenotetext{e}{$f=2.975\times10^{7}$.}



\end{deluxetable*}
\endgroup